\documentclass[11pt]{article}
\usepackage{amsfonts,amsbsy,eucal}

\newsymbol\Bbbk 207C

\begin{document}

%
%
\catcode`\@=11 

%
\global\newcount\secno \global\secno=0
\global\newcount\meqno \global\meqno=1

\def\newsec#1{\global\advance\secno by1
\global\subsecno=0\eqnres@t
\section{#1}}
\def\eqnres@t{\xdef\secsym{\the\secno.}\global\meqno=1}
\def\sequentialequations{\def\eqnres@t{\bigbreak}}\xdef\secsym{}
\global\newcount\subsecno \global\subsecno=0
\def\subsec#1{\global\advance\subsecno by1
\subsection{#1}}
\def\subsubsec#1{\subsubsection{#1}}

\def\draftmode{\message{ DRAFTMODE }
\writelabels
{\count255=\time\divide\count255 by 60 \xdef\hourmin{\number\count255}
\multiply\count255 by-60\advance\count255 by\time
\xdef\hourmin{\hourmin:\ifnum\count255<10 0\fi\the\count255}}}
\def\nolabels{\def\wrlabeL##1{}\def\eqlabeL##1{}\def\reflabeL##1{}}
\def\writelabels{\def\wrlabeL##1{\leavevmode\vadjust{\rlap{\smash%
{\line{{\escapechar=` \hfill\rlap{\tt\hskip.03in\string##1}}}}}}}%
\def\eqlabeL##1{{\escapechar-1\rlap{\tt\hskip.05in\string##1}}}%
\def\reflabeL##1{\noexpand\llap{\noexpand\sevenrm\string\string\string##1}
}}

\nolabels

\def\eqn#1#2{
\xdef #1{(\secsym\the\meqno)}
\global\advance\meqno by1
$$#2\eqno#1\eqlabeL#1
$$}

\def\eqalign#1{\null\,\vcenter{\openup\jot\m@th
\ialign{\strut\hfil$\displaystyle{##}$&$\displaystyle{{}##}$\hfil
\crcr#1\crcr}}\,}

\def\foot#1{\footnote{#1}}

%
\global\newcount\refno \global\refno=1
\newwrite\rfile
\def\ref{[\the\refno]\nref}
\def\nref#1{\xdef#1{[\the\refno]}
\ifnum\refno=1\immediate\openout\rfile=refs.tmp\fi
\global\advance\refno by1\chardef\wfile=\rfile\immediate
\write\rfile{\noexpand\bibitem{\string#1}}\findarg}
\def\findarg#1#{\begingroup\obeylines\newlinechar=`\^^M\pass@rg}
{\obeylines\gdef\pass@rg#1{\writ@line\relax #1^^M\hbox{}^^M}%
\gdef\writ@line#1^^M{\expandafter\toks0\expandafter{\striprel@x #1}%
\edef\next{\the\toks0}\ifx\next\em@rk\let\next=\endgroup\else\ifx\next\empty

\else\immediate\write\wfile{\the\toks0}\fi\let\next=\writ@line\fi\next\relax

}}
\def\striprel@x#1{} \def\em@rk{\hbox{}}
\def\lref{\begingroup\obeylines\lr@f}
\def\lr@f#1#2{\gdef#1{\ref#1{#2}}\endgroup\unskip}
\def\semi{;\hfil\break}
\def\addref#1{\immediate\write\rfile{\noexpand\item{}#1}}
\def\listrefs{

}
\def\startrefs#1{\immediate\openout\rfile=refs.tmp\refno=#1}
\def\xref{\expandafter\xr@f}\def\xr@f[#1]{#1}
\def\refs#1{\count255=1[\r@fs #1{\hbox{}}]}
\def\r@fs#1{\ifx\und@fined#1\message{reflabel \string#1 is
undefined.}%
\nref#1{need to supply reference \string#1.}\fi%
\vphantom{\hphantom{#1}}\edef\next{#1}\ifx\next\em@rk\def\next{}%
\else\ifx\next#1\ifodd\count255\relax\xref#1\count255=0\fi%
\else#1\count255=1\fi\let\next=\r@fs\fi\next}
\newwrite\lfile
{\escapechar-1\xdef\pctsign{\string\%}\xdef\leftbracket{\string\{}
\xdef\rightbracket{\string\}}\xdef\numbersign{\string\#}}
\def\writedefs{\immediate\openout\lfile=labeldefs.tmp
\def\writedef##1{%
\immediate\write\lfile{\string\def\string##1\rightbracket}}}
\def\writestop{\def\writestoppt{\immediate\write\lfile{\string\pageno%
\the\pageno\string\startrefs\leftbracket\the\refno\rightbracket%
\string\def\string\secsym\leftbracket\secsym\rightbracket%
\string\secno\the\secno\string\meqno\the\meqno}\immediate\closeout\lfile}}

\def\writestoppt{}\def\writedef#1{}

\catcode`\@=12 
%

%
\def\noblackbox{\overfullrule=0pt}
\hyphenation{anom-aly anom-alies coun-ter-term coun-ter-terms
}
\def\inv{^{\raise.15ex\hbox{${\scriptscriptstyle -}$}\kern-.05em 1}}
\def\dup{^{\vphantom{1}}}
\def\Dsl{\,\raise.15ex\hbox{/}\mkern-13.5mu D}
\def\dsl{\raise.15ex\hbox{/}\kern-.57em\partial}
\def\del{\partial}
\def\Psl{\dsl}
\def\tr{{\rm tr}} \def\Tr{{\rm Tr}}
\font\bigit=cmti10 scaled \magstep1
\def\biglie{\hbox{\bigit\$}} 
\def\lspace{\ifx\answ\bigans{}\else\qquad\fi}
\def\lbspace{\ifx\answ\bigans{}\else\hskip-.2in\fi} 
\def\boxeqn#1{\vcenter{\vbox{\hrule\hbox{\vrule\kern3pt\vbox{\kern3pt
\hbox{${\displaystyle #1}$}\kern3pt}\kern3pt\vrule}\hrule}}}
\def\tilde{\widetilde} 
\def\hat{\widehat}
%
\def\e#1{{\rm e}^{^{\textstyle#1}}}
\def\grad#1{\,\nabla\!_{{#1}}\,}
\def\gradgrad#1#2{\,\nabla\!_{{#1}}\nabla\!_{{#2}}\,}
\def\ph{\varphi}
\def\psibar{\overline\psi}
\def\om#1#2{\omega^{#1}{}_{#2}}
\def\vev#1{\langle #1 \rangle}
\def\lform{\hbox{$\sqcup$}\llap{\hbox{$\sqcap$}}}
\def\darr#1{\raise1.5ex\hbox{$\leftrightarrow$}\mkern-16.5mu #1}
\def\lie{\hbox{\it\$}} 
\def\ha{{1\over2}}
\def\half{{\textstyle{1\over2}}} 
\def\roughly#1{\raise.3ex\hbox{$#1$\kern-.75em\lower1ex\hbox{$\sim$}}}

%


\font\tencmmib=cmmib10  \skewchar\tencmmib='177
\font\sevencmmib=cmmib7 \skewchar\sevencmmib='177
\font\fivecmmib=cmmib5 \skewchar\fivecmmib='177
\newfam\cmmibfam
\textfont\cmmibfam=\tencmmib
\scriptfont\cmmibfam=\sevencmmib
\scriptscriptfont\cmmibfam=\fivecmmib
\def\cmmib#1{{\fam\cmmibfam\relax#1}}

\def\a{\alpha}
\def\b{\beta}
\def\c{\chi}
\def\d{\delta}  \def\D{\Delta}
\def\e{\varepsilon} \def\ep{\epsilon}
\def\f{\phi}  \def\F{\Phi}
\def\g{\gamma}  \def\G{\Gamma}
\def\k{\kappa}
\def\l{\lambda}  \def\La{\Lambda}
\def\m{\mu}
\def\n{\nu}
\def\r{\rho}
\def\vr{\varrho}
\def\o{\omega}  \def\O{\Omega}
\def\p{\psi}  \def\P{\Psi}
\def\s{\sigma}  \def\S{\Sigma}
\def\th{\theta}  \def\vt{\vartheta}
\def\t{\tau}
\def\w{\varphi}
\def\x{\xi}
\def\z{\zeta}
\def\U{\Upsilon}
\def\CA{{\cal A}}
\def\CB{{\cal B}}
\def\CC{{\cal C}}
\def\CD{{\cal D}}
\def\CE{{\cal E}}
\def\CF{{\cal F}}
\def\CG{{\cal G}}
\def\CH{{\cal H}}
\def\CI{{\cal I}}
\def\CJ{{\cal J}}
\def\CK{{\cal K}}
\def\CL{{\CMcal L}}
\def\CM{{\cal M}}
\def\CN{{\cal N}}
\def\CO{{\cal O}}
\def\CP{{\cal P}}
\def\CQ{{\cal Q}}
\def\CR{{\cal R}}
\def\CS{{\cal S}}
\def\CT{{\CMcal T}}
\def\CU{{\cal U}}
\def\CV{{\cal V}}
\def\CW{{\cal W}}
\def\CX{{\cal X}}
\def\CY{{\cal Y}}
\def\CZ{{\cal Z}}
\def\EJ{\mathfrak{J}}
\def\EM{\mathfrak{M}}
\def\ES{\mathfrak{S}}

\def\V{\mathbb{V}}
\def\E{\mathbb{E}}
\def\R{\mathbb{R}}
\def\C{\mathbb{C}}
\def\Z{\mathbb{Z}}
\def\A{\mathbb{A}}
\def\T{\mathbb{T}}
\def\L{\mathbb{L}}
\def\D{\mathbb{D}}
\def\Q{\mathbb{Q}}

\def\rd{\partial}
\def\grad#1{\,\nabla\!_{{#1}}\,}
\def\gradd#1#2{\,\nabla\!_{{#1}}\nabla\!_{{#2}}\,}
\def\om#1#2{\omega^{#1}{}_{#2}}
\def\vev#1{\langle #1 \rangle}
\def\darr#1{\raise1.5ex\hbox{$\leftrightarrow$}
\mkern-16.5mu #1}
\def\Ha{{1\over2}}
\def\ha{{\textstyle{1\over2}}}
\def\fr#1#2{{\textstyle{#1\over#2}}}
\def\Fr#1#2{{#1\over#2}}
\def\rf#1{\fr{\rd}{\rd #1}}
\def\rF#1{\Fr{\rd}{\rd #1}}
\def\df#1{\fr{\d}{\d #1}}
\def\dF#1{\Fr{\d}{\d #1}}
\def\DDF#1#2#3{\Fr{\d^2 #1}{\d #2\d #3}}
\def\DDDF#1#2#3#4{\Fr{\d^3 #1}{\d #2\d #3\d #4}}
\def\ddF#1#2#3{\Fr{\d^n#1}{\d#2\cdots\d#3}}
\def\fs#1{#1\!\!\!/\,}   
\def\Fs#1{#1\!\!\!\!/\,} 
\def\roughly#1{\raise.3ex\hbox{$#1$\kern-.75em
\lower1ex\hbox{$\sim$}}}
\def\ato#1{{\buildrel #1\over\longrightarrow}}
\def\up#1#2{{\buildrel #1\over #2}}
\def\opname#1{\mathop{\kern0pt{\rm #1}}\nolimits}
\def\tr{\opname{Tr}}
\def\Re{\opname{Re}}
\def\Im{\opname{Im}}
\def\End{\opname{End}}
\def\dim{\opname{dim}}
\def\vol{\opname{vol}}
\def\group#1{\opname{#1}}
\def\SU{\group{SU}}
\def\U{\group{U}}
\def\SO{\group{SO}}
\def\pr{\prime}
\def\ppr{{\prime\prime}}
\def\bs{\mathib{s}}
\def\bbs{\bar\mathib{s}}
\def\Dp{\rd_{\!A}}
\def\Dpp{\bar\rd_{\!A}}
\def\Da{d_{\!A}}
\def\bari{\bar\imath}
\def\barj{\bar\jmath}
\def\mapr#1{\!\smash{\mathop{\longrightarrow}\limits^{#1}}\!}
\def\mapl#1{\!\smash{\mathop{\longleftarrow}\limits^{#1}}\!}
\def\mapbr{\!\smash{\mathop{\longrightarrow}\limits^{\bbs_+}}\!}
\def\mapbl{\!\smash{\mathop{\longleftarrow}\limits^{\bbs_-}}\!}
\def\mapd#1{\Big\downarrow\rlap{$\vcenter{#1}$}}
\def\mapu#1{\Big\uparrow\rlap{$\vcenter{#1}$}}
\def\maprd{\rlap{\lower.3ex\hbox{$\scriptstyle\bs_+$}}\searrow}
\def\mapld{\swarrow\!\!\!\rlap{\lower.3ex\hbox{$\scriptstyle\bs_-$}}}
\def\ne{\nearrow}
\def\se{\searrow}
\def\nw{\nwarrow}
\def\sw{\swarrow}
\def\etal{et al.}
\def\git{/\kern-.25em/}
\def\Ker{\hbox{Ker}\;}


\def\lin#1{\begin{flushleft}
{\sc #1}
\end{flushleft}
}
\def\linn#1{\noindent $\bullet$ {\it #1} \par}

\makeatother




\def\etag#1{\eqnn#1\eqno#1}
\def\subsubsec#1{{\medbreak\smallskip\noindent{\it #1 } }}
\def\bos#1{\boldsymbol{#1}}
\def\tQ{\tilde{\bos{Q}}}
\def\inbar{\,\vrule height1.5ex width.4pt depth0pt}
\def\IC{\relax{\hbox{$\inbar\kern-.3em{\rm C}$}}}
\def\IR{\relax{\rm I\kern-.18em R}}
\font\cmss=cmss10 \font\cmsss=cmss10 at 7pt
\def\IZ{\relax\ifmmode\mathchoice
{\hbox{\cmss Z\kern-.4em Z}}{\hbox{\cmss Z\kern-.4em Z}}
{\lower.9pt\hbox{\cmsss Z\kern-.4em Z}}
{\lower1.2pt\hbox{\cmsss Z\kern-.4em Z}}\else{\cmss Z\kern-.4em Z}\fi}

\def\mq{{q}}
\def\mp{{p}}
\def\mH{\mathfrak{H}}
\def\mh{\mathfrak{h}}
\def\ma{\mathfrak{a}}
\def\ms{\mathfrak{s}}
\def\mm{\mathfrak{m}}
\def\mn{\mathfrak{n}}

\def\Hoch{{\tt Hoch}}
\def\mt{\mathfrak{t}}
\def\ml{\mathfrak{l}}
\def\mT{\mathfrak{T}}
\def\mL{\mathfrak{L}}
\def\mg{\mathfrak{g}}
\def\mf{\mathfrak{f}}
\def\mh{\mathfrak{h}}
\def\md{\mathfrak{d}}





\centerline{\LARGE \sc
Pursuing The Quantum World
}
\smallskip
\centerline{\large \sc Flat Family Of QFTs
And Quantization Of $d$-Algebras}
\bigskip
\centerline{Jae-Suk Park}
\medskip
\centerline{Department of Mathematics, POSTECH,  Pohang, 790-784, Korea}
\bigskip
\centerline{\it Dedicated To the Memory of Youngjai Kiem}

\begin{quote}
\footnotesize
Exploiting the path integral approach al la Batalin and Vilkovisky,
we show that any anomaly-free Quantum Field Theory (QFT) comes
with a family parametrized by certain moduli space $\mathfrak{M}$, which tangent
space at the point corresponding to the initial QFT is given by
the space of all observables. Furthermore the tangent bundle over $\mathfrak{M}$
is equipped with flat quantum connection, which can be used to determine
all correlation functions of the family of QFTs. We also argue that
considering family of QFTs is an inevitable step, due to the fact that
the products of quantum observables are not quantum observables
in general,  which leads to a new  "global" perspective on quantum world.
We also uncover structure of $d$-algebra in the large class of
$d$-dimensional QFT. This leads to an universal quantization machine
for $d$-algebras decorated by algebro-differential-topology of
$(d+1)$-manifolds as well as a new perspective on differential-topology
of low dimensions. This paper is a summary of a forthcoming paper
of this author.

\end{quote}

\newsec{Prelude}

The physical reality is supposed to be genuinely quantum
such that anything classical, if any, is just an approximated or derived notion.
On the other hand, classical physical theory, historically, came first and, then,
certain procedure called quantization has been adapted to  mimic quantum description.
This idea of quantization of classical object or notion is obviously limited, though
it gives us an useful guide of our journey toward genuine quantum world, 
if we take educated
steps  with constant renormalization of our viewpoints - namely by pursuing
clues left out from quantum world.  
Only in "the" end, if any, of such journey we may have proper definition 
of what it meant to be Quantum.  It seems also reasonable to assume
that  mathematical structures associated with classical physics
as shadows of certain quantum mathematics. 
This, in particular, implies that essential properties of supposedly
classical world may be understood naturally in the quantum 
perspective. 

This article is a 
summary of  this author's pursuit to understand  Quantum Fields Theory (QFT)
for the last $3$ years, which details shall appear elsewhere \cite{P2}.

We shall begin with, in section $2$, a  reflection  on  the idea of quantization in 
the path integral approach a la Batalin-Vilkovisky (BV) formalism, which will lead, in section $3$, to notion of quantum flat structure on a family of  QFTs. 
The story goes as follows; In the beginning  we may start from certain
classical field theory characterized by a classical action functional with certain (gauge) symmetry. The BV quantization scheme, then, 
suggests to build up, out of the given classical structure, 
a mathematical structure dubbed as {\it quantum weakly
homotopy Lie $(-1)$-algebroid}, which  procedure is controlled
by BV algebra  - the odd nilpotent and order $2$ BV operator $\bos{\Delta}$ 
together ordinary product. A crucial property of the BV operator
$\bos{\Delta}$ is that it is not a derivation of the product which
leads to a problem that the products of quantum observables
of a given QFT are not quantum observables in general. Then we argue that
the above problem can be resolved by considering family of
QFTs parametrized certain moduli space $\mathfrak{M}$ 
defined by set of equivalence classes of solutions of the celebrated
quantum master equation. Here the initial QFT is interpreted as
a base point $o$ in $\mathfrak{M}$ such that certain basis of quantum
observables is regarded as a basis of tangent space to $\mathfrak{M}$
at the base point $o$. Now the problem involving products of quantum observables
shall be resolved by reaching out to certain formal (beyond infinitesimal)
neighborhood of $o$ in $\mathfrak{M}$, introducing certain quantum products
of quantum observables.  Then we shall have a system of differential
equations satisfied by generating functional of path integrals for the family of QFTs.

All of these may be summarized by {\it quantum flat
structure}, which roughly suggest to build up a kind of formal graded
bundle $\mathfrak{Q}\rightarrow \mathfrak{M}$ 
equipped with certain graded flat connection in formal power series of the
Planck constant $\hbar$ such that path integrals can be described
as flat sections. This structure seems to suggest to find certain
completion $\overline\mathfrak{Q}\rightarrow \overline\mathfrak{M}$
including various degenerated  limits of $\mathfrak{Q}\rightarrow \mathfrak{M}$,
where various perturbative QFTs may corresponds to singular points.
Unfortunately this author does not have a good understanding on the
above issues.

The section $4$ is a sketch of a program to understand a large class of QFTs
on a $(d+1)$-dimensional manifold with or without boundaries. 
This section is essentially an elaboration and generalization of this author's
previous work. Some reflection on $(d+1)$-dimensional field theory 
a la BV quantization scheme suggests that one may associate pre-QFT
on smooth oriented $(d+1)$-manifold with any structure of symplectic
$(d+1)$-algebra. On $(d+1)$-manifold with boundary the BV quantization
scheme suggest that one can associate a structure of strongly homotopy
Lie $d$-algebroid on the boundary. This suggest application of
$(d+1)$-dimensional QFT with boundary to deformation theory of so-called
$d$-algebra \cite{K2}, generalizing the deformation quantization story 
of Kontsevich \cite{K1}. This section is an elaboration and generalization
of this author's previous work \cite{P1}, where we assumed much narrow scope. 
We shall
also introduce possible notion of quantum cobordism.

The pre-QFT constructed from a symplectic $s$-algebra would become
actual QFT if the set of equivalence class of solutions of Maurer-Cartan
equation is isomorphic  to the moduli space $\mathfrak{M}$ discussed
in Section $3$, which notion requires proper definition of the BV operator 
$\bos{\Delta}$. It is unfortunate that this author does not have good understanding
in a proper definition of $\bos{\Delta}$, which notion is the nerve of
quantum flat structure.  Assuming that we have a good definition
of $\bos{\Delta}$ our model shall not only provided an universal (?)
quantization machine of $d$-algebras as well as new arena of
differential-topological invariants of low dimensional manifolds.

For the purpose this article we didn't present any explicit
example adopting the general program. This author, however,
has been implanted the idea to produce many of known QFTs
and numerous new examples for the last $3$-years, which 
already appeared or shall appear in some future publications \cite{P2,P3,PR,PT,HP,HP1,HP3,OP1,OP2}.
Throughout this paper we shall restrict to the case, for the sake of simplicity, that a field
with even ghost number $U$ is commuting, while a field with
odd ghost number $U$ is anti-commuting under the ordinary product.
This restriction means, in particular, that an object with
ghost number $U=0$ is commuting and, thus, excluding physical
fermions. It is straight forward to generalize the results in this paper
including physical fermions.

This author has chosen a rather vague but symbolic title of this article
with an intention. 

I would like to dedicate this small work to the memory of Youngjai Kiem.
He was a very good friend of mine and a talented young physicist who passed away 
by a tragic accident. His brief but enthusiastic life
had been devoted to, in my humble opinion, pursuing quantum world
besides from his late family and friends. 
In the last semester of his life he also kindly provided me
a visiting position in KAIST with an excellent research
environment.

\newsec{From Classical Field Theory and Its Symmetry}

We may view the history of understanding QFT as a long journey toward 
to genuine quantum world starting from a small corner near a classical world,
which may be described in terms of certain classical field theory
defined by an classical action functional $\bos{s}$ with certain  gauge
symmetry. In the quantum theory the central object is not the
action functional but Feynman Path Integral \cite{F}. The gauge symmetry
of classical action functional requires additional fields called
Faddev-Popov (FP) ghosts in the path integral \cite{FP}, which 
replace the gauge symmetry to odd global symmetry generated
by BRST (Becchi-Rouet-Stora and Tuytin) charge $\bos{q}$, i.e. $\bos{qs}=0$ \cite{BRS,T}, 
which satisfies $\bos{q}^2=0$, in general, modulo classical 
equations of motion of $\bos{s}$. We can say that the above was the
first step toward the world of QFT. 

To be brief  we begin with the state of arts scheme of quantization
pioneered by Batalin-Vilkovisky  \cite{BV}, see also
\cite{TNP,S,W2,W3,Z,Getzler,AKSZ,St2,K1,P1} for its various
aspects, and look back the history
as a revisionist.

\subsection{The BV Quantization I}

\def\insum{\int\!\!\!\!\!\!\!\!\!\sum}

Let $\{\phi^A\}$ denotes collectively all classical fields, ghosts, anti-ghost multiplets
etc etc, which may be called {\tt fields}. Here the indices $\{A\}$ are understood
as both continuous and discrete, or we may say $\{\phi^A\}$
is certain coordinates on an infinite dimensional graded space $\bos{\CL}$,
where the grading are specified by an integral ghost number $U \in \Z$
and an $\Z_2$ grading called parity .
BV introduced
so called set of {\tt anti-fields} $\{\phi^\bullet_A\}$ for each {\tt fields} such that
\eqn\aaa{
U(\phi^A) + U(\phi^\bullet_A) = -1,\qquad
|\phi^A| + |\phi^\bullet_A| = 1,
}
where the notation $|\star|$ means the parity (equivalently the statistics) $0$ (even) or $1$ (odd)
of the expression $\star$ such that objects with the even parity are commuting
while objects with the odd parity are anti-commuting.
The space of all {\tt fields} and {\tt anti-fields} can be viewed as
the total space $\bos{\CT}\simeq T^*[-1]\bos{\CL}$ of twisted by $[-1]$ cotangent
bundle to $\bos{\CL}$ - twisting by $[-1]$ simply means the convention
\aaa\ of ghost numbers.\foot{In general  $T^*[s]\bos{\CL}$
shall means the conventions $U(\phi^A) +U(\phi^\bullet_A)=s$
and $|\phi^A| +|\phi^\bullet_A| = s\hbox{ mod } 2$.}
Then we have a canonical odd symplectic
structure $\bos{\o}$ of ghost number $U=-1$ on $\bos{\CT}$, i.e., $U(\bos{\o})=-1$
and $|\bos{\o}|=1$.
We shall restrict to the case, for the sake of simplicity, that a field
with even (odd) ghost number $U$  is even (odd) parity.

Let $\mT[[\hbar]]$ be the space of functions on 
$\bos{\CT}$ formal power series in Planck constant $\hbar$ with $U(\hbar)=0$,
which  is also a graded space by the ghost number $U$;
\eqn\aab{
\mT[[\hbar]]=\bigoplus_{k\in \Z} \mT[[\hbar]]_k.
}
Note that we have ordinary product $\cdot$ carrying $U=0$, 
which is graded commutative
and graded associative, i.e., $\CO_1\cdot \CO_2 = (-1)^{|\CO_1|\cdot|\CO_2|}\CO_2\cdot
\CO_1$ where
$\CO_1,\CO_2 \in \mT[[\hbar]]$.

BV introduced an order $2$  odd differential $\bos{\Delta}$
operator  carrying $U=1$
\eqn\aac{
\bos{\Delta} "\!=\!" (-1)^{|\phi^A|+1} 
\Fr{\d_r^2}{\d \phi^A \d \phi^\bullet_A},
\qquad \bos{\Delta}: \mT[[\hbar]]_k \longrightarrow \mT[[\hbar]]_{k+1},
}
satisfying
\eqn\aad{
\bos{\Delta}^2=0.
}
We denote
$\Fr{\d_r}{\d\phi}$ and $\Fr{\d_l}{\d\phi}$ the right and left differentiation,
respectively. We shall also use convention that repeated up and down indices 
are summed (or integrated) over.

A {\it BV action functional} $\bos{S} \in \mT[[\hbar]]_{0}$
is an even function on $\bos{\CT}$ in formal power series of
$\hbar$ satisfying the celebrated {\it master equation}
\eqn\master{
\bos{\Delta} e^{-\bos{S}/\hbar} =0.
}
We should not forget to emphasis that
the definition of $\bos{\Delta}$ is formal or even symbolic
and always requires careful regularization to ensure the
crucial property  $\bos{\Delta}^2=0$. We shall not care about
it at this stage and simply assume that the space $\bos{\CT}$ is
equipped with such an operator - but
QFT is largely characterized by  $\bos{\Delta}$.

Being an order $2$ differential $\bos{\Delta}$ is not a derivation
of the product;
\eqn\aaf{
(-1)^{|\CO_1|}\bos{\Delta}(\CO_1\cdot\CO_2) 
-  (-1)^{|\CO_1|}\bos{\Delta} \CO_1 \cdot \CO_2 - 
\CO_1\cdot \bos{\Delta} \CO_2
= \left(\CO_1,\CO_2\right),
}
where the  bracket $(\star,\star)$, called BV bracket,
is identical to the odd Poisson bracket carrying $U=1$
associated with the symplectic
structure $\bos{\o}$ of ghost number $U=-1$ on $\bos{\CT}$.;
\eqn\aag{
\left(\star,\star\right): \mT[[\hbar]]_{k_1}\otimes\mT[[\hbar]]_{k_2}
\longrightarrow \mT[[\hbar]]_{k_1+k_2 +1}.
}
The BV bracket operation
may be expressed as
\eqn\aah{
\left(\CO_1,\CO_2\right) = 
\left( \Fr{\d_r \CO_1}{\d \phi^A}\cdot\Fr{\d_l\CO_2}{\rd \phi^\bullet_A}
 -\Fr{\d_r \CO_1}{\rd \phi^\bullet_A}\cdot\Fr{\d_l\CO_2}{\rd \phi^A}
\right)
}
and satisfies
\eqn\aaha{
\eqalign{
\left(\CO_1,\CO_2\right) &=- (-1)^{(|\CO_1|+1)(|\CO_2|+1)}\left(\CO_2,\CO_1\right),\cr
\left(\CO_1,\CO_2\cdot\CO_3\right) &= \left(\CO_1,\CO_2\right)\cdot\CO_3
+(-1)^{(|\CO_1|-1)|\CO_2|}\CO_2\cdot\left(\CO_1,\CO_3\right),\cr
\left(\CO_1,(\CO_2,\CO_3)\right)&=\left(\left(\CO_1,\CO_2\right),\CO_3\right)
+(-1)^{(|\CO_1|-1)(|\CO_2|-1)}\left(\CO_2,\left(\CO_1,\CO_3\right)\right).
}
}
Now the master equation \master\ is equivalent to the
following equation
\eqn\aai{
-\hbar\bos{ \Delta S} + \Fr{1}{2}(\bos{S},\bos{S})=0.
}

Let's consider the odd Hamiltonian vector $\bos{Q_S}$ 
\eqn\aaj{
\bos{Q_S} = (\bos{S},\ldots),
}
carrying $U=1$ as well as  the following odd operator $\bos{K_S}$
\eqn\aak{
\bos{K_S} := -\hbar\bos{\Delta} + \bos{Q_S},\qquad
\bos{K_S}: \mT[[\hbar]]_k \longrightarrow \mT[[\hbar]]_{k+1},
}
carrying $U=1$. 
The master equation \aai\  implies that
\eqn\aal{
\bos{K}^2_{\bos{S}}=0,
}
while $\bos{Q}^2_{\bos{S}}\neq 0$ in general.
Thus we have the BV complex
\eqn\aam{
\left(\bos{K_S}, \mT[[\hbar]] =\bigoplus_k \mT[[\hbar]]_k\right)
}
The set of {\it observables} of given QFT is identified with set of cohomology classes 
of the above BV complex, beautiful!
There is another crucial property, which looks vexing initially
but beautiful, that the products $\bos{K_S}$-closed elements
in $\mT[[\hbar]]$ are not $\bos{K_S}$-closed in general, 
since $\bos{\Delta}$ is not a derivation of products.
We note that both $\bos{Q_S}$ and $\bos{\Delta}$ are
derivations of the BV bracket.

At this stage we trace back to our starting point in the following
two subsections.

\subsection{Quantum Weakly Homotopy Lie $(-1)$-Algebroid}

We consider a solution of master as the formal power series in $\hbar$
as
\eqn\baa{
\bos{S} = \bos{S}^{(0)} + \sum_{\ell=1}^\infty \hbar^\ell \bos{S}^{(\ell)},
\qquad
\bos{\Delta} \bos{S}^{(\ell)} = \Fr{1}{2}\sum_{r+s=\ell+1}\left(\bos{S}^{(r)} , \bos{S}^{(s)}\right)
\hbox{ for }\forall n \geq 0
}
such that
\eqn\bab{
\eqalign{
0&=\left(\bos{S}^{(0)},\bos{S}^{(0)}\right),\cr
\bos{\Delta S}^{(0)} &= \left(\bos{S}^{(0)} , \bos{S}^{(1)}\right),\cr
}
}
etc. etc.
Then one may Taylor expand the each term $\bos{S}^{(\ell)}$ around
the space $\bos{\CL}$ of {\tt fields} 
\eqn\bac{
\eqalign{
\bos{S}^{(\ell)} &= \sum_{n=0}^\infty \bos{M}^{(\ell)}_n,\cr
\bos{M}^{(\ell)}_n & 
= \insum (\bos{m}(\phi)^{(\ell)})^{A_1\ldots A_n} \phi^\bullet_{A_1}\ldots  \phi^\bullet_{A_n}
}
}
leading to double infinite sequence of relations
\eqn\bad{
\bos{\Delta} \bos{M}^{(\ell)}_n = \Fr{1}{2}\sum_{r+s=\ell+1}\sum_{p+q =n}\left(\bos{M}^{(r)}_p , \bos{M}^{(s)}_q\right).
}
For each $\bos{M}^{(\ell)}_n$ one may assign $n$-poly differential
operator $\bos{m}^{(\ell)}_n$ acting on the $n$-th tensor product $\mL^{\otimes m}$ of the space $\mL$ of functions
on $\bos{\CL}$ such as; 
\eqn\bae{
\eqalign{
\bos{m}^{(\ell)}_0:  &\bos{\CL} \rightarrow \Bbbk,\cr
\bos{m}^{(\ell)}_n: &\mL^{\otimes n}\rightarrow \mL,\qquad\hbox{for } n\geq 1
} 
}
by canonically "quantization", i.e., replacing
the BV bracket $(\phi^A, \phi^\bullet_B)=\d^{A}{}_B$
to commutators of operators  $[\hat \phi^A, \hat\phi^\bullet_B]=\d^{A}{}_B$;
naively $\hat\phi^B = \phi^B$ and $\hat\phi^\bullet_A = \Fr{\d}{\d \phi^A}$.
In this way we have following double infinite sequence
\eqn\baf{
\matrix{
\bos{s} & \bos{q} &\bos{m}^{(0)}_2 &\bos{m}^{(0)}_3 & \cdots\cr
\bos{m}^{(1)}_0 & \bos{m}^{(1)}_1 &\bos{m}^{(1)}_2 &\bos{m}^{(1)}_3 & \cdots\cr
\bos{m}^{(2)}_0 & \bos{m}^{(2)}_1 &\bos{m}^{(2)}_2 &\bos{m}^{(2)}_3 & \cdots\cr
\vdots &\vdots &\vdots& \vdots &\cdots\cr
}
}
where we denoted
\eqn\bafa{
\eqalign{
\bos{m}^{(0)}_0 &=\bos{s},\cr
\bos{m}^{(0)}_1 &=\bos{q}.\cr 
}
}
It is not difficult to check $\bos{m}^{(\ell)}_n$ carry ghost number
$U=n$ for $\forall \ell$. For instance $\bos{s}$ and $\bos{q}$
have the ghost number $U=0$ and $U=1$, respectively.

We consider the
classical master equation;
\eqn\tree{
\left(\bos{S}^{(0)},\bos{S}^{(0)}\right)=0,
} 
which can be decomposed as
$\sum_{p+q =m}\left(\bos{M}^{(0)}_p , \bos{M}^{(0)}_q\right)=0$
for $\forall n\geq 0$
, i.e.,
\eqn\bag{
\eqalign{
0&=-\left(\bos{M}^{(0)}_0 , \bos{M}^{(0)}_1\right),\cr
\Fr{1}{2}\left(\bos{M}^{(0)}_1 , \bos{M}^{(0)}_1\right)&=-\left(\bos{M}^{(0)}_0 , \bos{M}^{(0)}_2\right),\cr
\left(\bos{M}^{(0)}_1 , \bos{M}^{(0)}_2\right)&=-\left(\bos{M}^{(0)}_0 , \bos{M}^{(0)}_3\right),\cr
\Fr{1}{2}\left(\bos{M}^{(0)}_2 , \bos{M}^{(0)}_2\right)+
\left(\bos{M}^{(0)}_1 , \bos{M}^{(0)}_3\right)&=-\left(\bos{M}^{(0)}_0 , \bos{M}^{(0)}_4\right),\cr
\vdots\quad\qquad &=\quad\qquad\vdots
}
}
We define a structure of {\it weakly homotopy Lie $(-1)$-algebroid}
on $\bos{\CL}$ by
the sequence 
\eqn\whs{
\bos{s}, \bos{q},\bos{m}^{(0)}_2,\bos{m}^{(0)}_3, \cdots
}
associated with a solution of the tree level master equation \bag.\foot{The 
terminology weakly homotopy 
Lie $(-1)$-algebroid is a composition from the notions
of (weakly or strongly) homotopy Lie algebra, $d$-algebra, and Lie algebroid.
Strongly homotopy algebra has been first introduced by Stasheff \cite{S1}.
For the notion of $d$-algebra with $d=2,3,\ldots$ see Kontsevich \cite{K2}.  
Lie algebroid may be regarded as the infinitesimal of Lie groupoid,
(see Weinstein for an introduction). The relation between
solution classical master equation \tree\ and homotopy algebra
seems to be trace back to Witten \cite{W3}, Zwiebach \cite{Z} and to 
Alexandrov et. al., \cite{AKSZ}.
}
In more familiar terminology $\bos{s}$,
which is a function on $\bos{\CL}$, is nothing but
the {\it classical action functional}  from which we may have been started.
We also note that
\eqn\baga{
\bos{s} =\lim_{\hbar\rightarrow 0}\bos{S}|_{\bos{\CL}}.
}
The first order differential operator $\bos{q}$ with
ghost number $U=1$ is the BRST operator, which equivalent to
an odd vector field on $\bos{\CL}$. 
The first relation in RHS
of \bag\ is the familiar BRST invariance of the classical action functional
\eqn\bah{
\bos{qs}=0.
}
The second relation in RHS of \bag\ implies that
nilpotency of the BRST operator  $\bos{q}$
may be violated up to the equation of motion of $\bos{s}$
\eqn\bak{
\bos{q}^2(anything)= -\bos{m}^{(0)}_2(\bos{s}, anything)
\propto
\d \bos{s}.
}
Thus BV quantization suggests to find the whole
sequence \whs\
of a structure of weakly homotopy Lie $(-1)$-algebroid from the
classical action functional $\bos{s}$ and its symmetry.
Such a procedure would produce the tree
level BV action functional $\bos{S}^{(0)}$ satisfying the first equation in \bab.
Then one should check if $\bos{\Delta S}^{(0)}=0$ and, otherwise, find
$\bos{S}^{(1)}$ satisfying the second equation in \bab, etc. etc.

We may also combine the sequence $(\bos{m}^{(0)}_n,\bos{m}^{(1)}_n,\bos{m}^{(2)}_n,
\cdots)$ as
\eqn\bafb{
\bos{m}_n = \bos{m}^{(0)}_n + \sum_{\ell=1}^\infty \hbar^\ell\bos{m}^{(\ell)}_n
}
such that
\eqn\bafc{
\eqalign{
\bos{m}_0:  &\bos{\CL} \rightarrow \Bbbk[[\hbar]],\cr
\bos{m}_n: &\mL[[\hbar]]^{\otimes n}\rightarrow \mL[[\hbar]],\qquad\hbox{for } n\geq 1
} 
}
We may call 
the sequence 
$(\bos{m}_0, \bos{m}_1,\bos{m}_2,\bos{m}_3, \cdots)$
a structure of {\it quantum weakly homotopy Lie $(-1)$-algebroid}
on $\bos{\CL}$, which classical limit $\hbar=0$ is
the structure $(\bos{s}, \bos{q},\bos{m}^{(0)}_2,\bos{m}^{(0)}_3, \cdots)$
of weakly homotopy Lie $(-1)$-algebroid on $\bos{\CL}$.
We may say  that construction of quantum weakly homotopy Lie $(-1)$-algebroid
is equivalent to defining a quantum field theory. Such a strategy has been
first adopted by Zwiebach in his construction of string field theory \cite{Z}. 
Note that,
for Zwiebach, non-vanishing $\bos{s}$ corresponds to non-conformal
background.

\subsection{Symmetry and  Anomaly: Resolutions and Obstructions}

\subsubsection{Gauge Symmetry and Resolutions}

Now we bring out an important issue, that we ignored so far,
which is actually related with the historical introduction of FP ghosts
and the notion of consistent anomaly.

Let's return to a solution $\bos{S}^{(0)}\in \mT_0$ to the classical master equation.
We know that another  solution $\bos{S}^{\pr(0)}\in \mT_0$ the classical master equation
gives rise to the equivalent physical theory if $\bos{S}^{\pr(0)}$
is related with $\bos{S}^{(0)}\in \mT_0$ by the ghost number and the parity preserving
canonical transformation. Such a canonical transformation
would be generated by an odd element $\bos{\Psi} \in \mT_{-1}$.
We also note that, since the BV bracket has $U=1$,
\eqn\msra{
\left(\star,\star\right): \mT_{-1}\otimes \mT_{-1}\longrightarrow \mT_{-1},
}
which means, together with the $1$st and the $3$rd relations in \aaha,
 that the BV bracket endows a structure of Lie algebra on
$\mT_{-1}$. Related to the above the following, ghost number preserving, adjoint action
by $\bos{\Psi} \in \mT_{-1}$
\eqn\msrb{
e^{ad_{\bos{\Psi}}}\circ (\star) = \star + \left(\bos{\Psi}, \star\right) + \Fr{1}{2!}\left(\bos{\Psi},\left(\bos{\Psi},\star\right)\right)
+\ldots
}
is equivalent to the canonical transformation.
Now we may define moduli space $\mathfrak{N}$
of classical BV action functional by
\eqn\msrc{
\mathfrak{N} =\left\{\bos{S}^{(0)}\in \mT_0| \left( \bos{S}^{(0)},\bos{S}^{(0)}\right)=0/\sim\right\}
}
where the equivalence $\sim$ is defined by the adjoint action by $\bos{\Psi} \in \mT_{-1}$.
Then a given QFT, modulo equivalence, 
with classical BV action functional $\bos{S}^{(0)}_o$,
corresponds to a point $o \in \mathfrak{N}$. 
Note that a tangent vector to $o\in \mathfrak{N}$ is 
$\Ker \bos{Q}_{\bos{S}^{(0)}_o}$ modulo the infinitesimal
version of the adjoint action \msrb.
It follows that the ghost number
$U=0$ part of $\bos{Q}_{\bos{S}^{(0)}_o}$-cohomology corresponds
to tangent space $T_o\mathfrak{N}$ to the point $o\in \mathfrak{N}$,
provided that  $\mathfrak{N}$ is smooth (around $o$).

We observe that
the adjoint action $ad_\Psi(\bos{S}^{(0)}_o)$ fixes 
$\bos{S}^{(0)}_o$
if $\Psi \in \Ker \bos{Q}_{\bos{S}^{(0)}_o} \cap \mT_{-1}$,
i.e,
\eqn\msrd{
\bos{Q}_{\bos{S}^{(0)}_o}\Psi\equiv \left(\bos{S}^{(0)}_o,\Psi\right)=0.
}
In other words the canonical transformation
generated by an element of  $\bos{Q}_{\bos{S}^{(0)}_o}$-cohomology $H^{-1}_{\bos{Q}_{\bos{S}^{(0)}_o}}$
in $\mT_{-1}$ is gauge symmetry of the classical BV action functional
$\bos{S}^{(0)}_o$.\foot{Note that the obvious
invariance generated by $\bos{\Psi} \in \Im  \bos{Q}_{\bos{S}^{(0)}_o} \cap \mT_{-1}$
is so called {\it fake symmetry}.}
Also the nontrivial $ \bos{Q}_{\bos{S}^{(0)}_o}$-cohomology
classes in $\mT_{-1}$ are related with sigularities the moduli space 
$\mathfrak{N}$ (around $o$), since the the correponding adjoint action \msrb\ fix
$\bos{S}^{(0)}_o$.

In our presentation so far we assumed implicitly that 
$H^{-1}_{\bos{Q}_{\bos{S}^{(0)}_o}}$ is trivial.
It is, however, certainly true that $H^{-1}_{\bos{Q}_{\bos{S}^{(0)}_o}}$ 
may be non-trivial in general.  It is, thus, more precise to say
that we assumed suitable trivialization or {\it resolution},
which notion shall be discussed briefly.

Let $\{\bos{\Upsilon}^{(0)}_a\}$ be a basis of $H^{-1}_{\bos{Q}_{\bos{S}^{(0)}_o}}$. 
Then we introduce dual basis $\{C^a\}$ with
ghost number $U=1$ and regard as a set of new {\tt fields} called {\it ghosts}.
We also introduce a set $\{C^\bullet_a\}$ of {\tt anti-fields} with
ghost number $U=-2$ for the ghosts $\{C^a\}$. Then extend
$\bos{\CT}$ to $\bos{\tilde\CT}=\bos{\CT}\times T^*[-1]
\left(H^{-1}_{\bos{Q}_{\bos{S}^{(0)}_o}}[1]\right)$
and the BV bracket $(\star,\star)$ to $\tilde{\left(\star,\star\right)}$.
We note that $\{\bos{\Upsilon}^{(0)}_a\}$ are, in general, functional of both the original
{\tt fields} and {\tt anti-fields} $(\phi^A,\phi^\bullet_A)$.
Since the BV bracket endows a structure of Lie algebra on $\mT_{-1}$
and since $\bos{Q}_{\bos{S}_o^{(0)}}$ is a derivation of the bracket,
we have a structure of Lie algebra on $H^{-1}_{\bos{Q}_{\bos{S}^{(0)}_o}}$
such that
\eqn\msre{
\left(\bos{\Upsilon}^{(0)}_a,\bos{\Upsilon}^{(0)}_b\right) = f_{ab}^c
\bos{\Upsilon}^{(0)}_c
\hbox{ mod } \Im  \bos{Q}_{\bos{S}_o^{(0)}},
}
with certain structure "constants" $f_{ab}^c=-f_{ba}^c$.\foot{Each component
of $\{f_{ab}^c\}$ can be even function, in general, with ghost number $U=0$.
The Jacobi identity of BV bracket suggest that
one may regard the pairts $(\{f_{ab}^c\},\{(\bos{\Upsilon}^{(0)}_a\})$
as structure of  Lie algebroid over graded space.
Remark  that we are talking about minimal model here such that
we also need to introduce {\it anti-ghost multiplets} for each ghost 
$C^a$ and their {\tt anti-fields}.
}
Then we may extend $\bos{S}_o^{(0)}$ to $\bos{\tilde S}_o^{(0)}$
as follows
\eqn\msrf{
\bos{\tilde S}_o^{(0)}=\bos{S}_o^{(0)} +  C^a _a \bos{\Upsilon}^{(0)}
+\Fr{1}{2}C^{a} C^{b} f_{ab}^c C^{\bullet}_{c},
}
which satisfies $\left(\bos{\tilde S}_o^{(0)},\bos{\tilde S}_o^{(0)}\right) =0$
from the relation \msre\ and its Jacobi identity as the BV bracket is
Lie bracket on $\mT_{-1}$.
Now we have odd nilpotent Hamiltonian vector field 
$\bos{Q}_{\bos{\tilde S}^{(0)}_o}=\left(\bos{\tilde S}_o^{(0)},\ldots\right)$
on  $\bos{\tilde\CT}$. Then we need to check if 
$H^{-1}_{\bos{\tilde Q}_{\bos{\tilde S}^{(0)}_o}}$ is trivial. Otherwise
we need to repeat the above procedure until we get trivial
$U=-1$ cohomology. 

We may call the above procedure small resolution 
$(\bos{\tilde\CT},\bos{\tilde S}_o^{(0)})$
of the pairs $(\bos{\CT},\bos{S}_o^{(0)})$.
Now we turn to the another source of obstruction in $\mathfrak{N}$.

\subsubsection{Constent Anomaly and Obstruction}

For a given solution $\bos{S}^{(0)}_o$ we consider nearby solution
$\bos{S}^{(0)}_o + \d \bos{S}^{(0)}$, where $\d \bos{S}^{(0)} \in H^0_{\bos{Q}_{\bos{S}^{(0)}_o}}$. It is obvious not every
$\d \bos{S}^{(0)}$ can be  the actual tangent vector of $\mathfrak{N}$
at $o$ since
$
\left(\d \bos{S}^{(0)},\d \bos{S}^{(0)}\right) \neq 0$,
in general. Let $\{\bos{O}^{(0)}_a\}$ be a basis of $H^0_{\bos{Q}_{\bos{S}^{(0)}_o}}$
and set $\d \bos{S}^{(0)}=t^a \bos{O}^{(0)}_a$, where $\{t^a\}$ be the dual
basis with $\{|t^a|\}=\{0\}$. We note that we may have
\eqn\msri{
\left(\bos{O}^{(0)}_a ,\bos{O}^{(0)}_b\right) = c_{ab}^c\bos{O}^{(0)}_{1c} 
+  \bos{Q}_{\bos{S}^{(0)}_o}\bos{O}^{(0)}_{ab}  \in \Ker \bos{Q}_{\bos{S}^{(0)}_o}\cap \mT_{1}
}
where $\bos{O}^{(0)}_{1c} \in H^1_{\bos{Q}_{\bos{S}^{(0)}_o}}$.
It is obvious that  there are no second order (in $\{t^a\}$)
corrections to $\bos{S}^{(0)}_o + t^a \bos{O}^{(0)}_a$ to satisfy
the classical master equation modulo the third order terms in 
$\{t^a\}$ unless $\left(\bos{O}^{(0)}_a ,\bos{O}^{(0)}_b\right)$ are  
$\bos{Q}_{\bos{S}^{(0)}_o}$-exact. Thus  $H^1_{\bos{Q}_{\bos{S}^{(0)}_o}}
\cap \left(H^0_{\bos{Q}_{\bos{S}^{(0)}_o}},H^0_{\bos{Q}_{\bos{S}^{(0)}_o}}\right)$
is obstruction of $\mathfrak{N}$.

Let's assume, temporarily, that $H^1_{\bos{Q}_{\bos{S}^{(0)}_o}}$ is
trivial. Then $\bos{S}^{(0)}_o + t^a \bos{O}^{(0)}_a +\Fr{1}{2}t^a t^b \bos{O}^{(0)}_{ab}$
solves the classical master equation modulo  $3$rd order terms in $\{t^a\}$;
\eqn\msrg{
t^a t^b t^c \left(\bos{O}^{(0)}_a, \bos{O}^{(0)}_{bc}\right).
}
To find the next order the
above term should be $\bos{Q}_{\bos{S}^{(0)}_o}$-exact.
It is suffice to show that the above is $\bos{Q}_{\bos{S}^{(0)}_o}$-closed
as $H^1_{\bos{Q}_{\bos{S}^{(0)}_o}}$ is
trivial. Note that
\eqn\msrh{
\bos{Q}_{\bos{S}^{(0)}_o}\left(\bos{O}^{(0)}_a, \bos{O}^{(0)}_{bc}\right)
=\left(\bos{O}^{(0)}_a,\bos{Q}_{\bos{S}^{(0)}_o} \bos{O}^{(0)}_{bc}\right)
=\left(\bos{O}^{(0)}_a ,\left(\bos{O}^{(0)}_b ,\bos{O}^{(0)}_c\right) \right)
}
It follows that
\eqn\msrj{
t^a t^b t^c\bos{Q}_{\bos{S}^{(0)}_o}\left(\bos{O}^{(0)}_a, \bos{O}^{(0)}_{bc}\right)
=t^a t^b t^c\left(\bos{O}^{(0)}_a ,\left(\bos{O}^{(0)}_b ,\bos{O}^{(0)}_c\right) \right)
=0,
}
where we used, in the last equality, the Jacobi-identity of BV bracket on $\mT_0\otimes\mT_0\otimes \mT_0$. Now by repeating the similar
procedure  iteratively one may establish the existence of solution
of classical master equation
in the form
\eqn\msrk{
\bos{S}^{(0)}(t)=\bos{S}^{(0)}_o +  t^a\bos{O}^{(0)}_a +\sum_{n=2}^\infty \Fr{1}{n!}t^{a_1}\ldots t^{a_n}
\bos{O}^{(0)}_{a_1\ldots a_n}.
}

We recall that $H^1_{\bos{Q}_{\bos{S}^{(0)}_o}}$ is known to be also related
with $1$-loop anomaly $\bos{\mathfrak{A}}^{(1)}_o$, defined in the BV language by
the formula;
\eqn\msrl{
\bos{\mathfrak{A}}^{(1)}_o = \bos{\Delta}\bos{S}^{(0)}_o -\bos{Q}_{\bos{S}^{(0)}_o}\bos{S}^{(1)}_o,
}
namely the $1$st obstruction for the classical BV action functional
$\bos{S}_o^{(0)}$ to be extended to quantum BV action functional (see \bab).
We note that
$\left(\bos{S}^{(0)}_o,\bos{\Delta S}^{(0)}_o\right)\equiv 
\bos{Q}_{\bos{S}^{(0)}_o} \bos{\Delta S}^{(0)}_o=0$,
since $\bos{\Delta}$ is a derivation of BV bracket and
$\left(\bos{S}^{(0)}_o,\bos{S}^{(0)}_o\right) =0$.
It follows, from \msrl, that 
\eqn\msrm{
\bos{Q}_{\bos{S}^{(0)}_o} \bos{\mathfrak{A}}^{(1)}_o=0,
}
which is equivalent to so called Wess-Zumino consistency condition
for anomaly \cite{WeZ}.
It follows that, if $H^1_{\bos{Q}_{\bos{S}^{(0)}_o}}$ is
trivial, the $1$-loop anomaly vanishes - $\bos{\mathfrak{A}}^{(1)}_o$ is a
$\bos{Q}_{\bos{S}^{(0)}_o}$-exact expression, if non-vanishing,
and we may simply modify $\bos{S}^{(1)}_o$ in \msrl.
It is shown by Troost et. al. \cite{TNP} that under the same condition the all loop anomaly
$\bos{\mathfrak{A}}_o = \sum_{\ell=1}^\infty \hbar^\ell\bos{\mathfrak{A}}^{(\ell)}_o$
also vanish, or, equivalently, for the given solution
$\bos{S}^{(0)}_o$ of classical master equation there exist
solution 
\eqn\msrma{
\bos{S}_o =\bos{S}^{(0)}_o +\sum_{\ell=1}^\infty\hbar^\ell \bos{S}^{(\ell)}_o,
}
of quantum master equation if  $H^1_{\bos{Q}_{\bos{S}^{(0)}_o}}$ is
trivial.

\subsubsection{From classical observables to quantum observables}

Here we  take a brief look  at the condition for $\bos{K_S}$-cohomology classes (that
of the BV complex \aam).
Consider
\eqn\bam{
\bos{Q_S} = (\bos{S},\ldots)=\sum_{n=0}^\infty\hbar^n \bos{Q}_{\bos{S}^{(n)}}=
\sum_{n=0}^\infty\hbar^n\left(\bos{S}^{(n)},\ldots\right),
}
so that the condition $\bos{K_S}\bos{O}$ has the following decompositions
\eqn\bana{
\bos{K_S}\bos{O} =0\quad \Longrightarrow
\left\{\eqalign{
 \bos{Q}_{\bos{S}^{(0)}}\bos{O}^{(0)}&=0,\cr
-\bos{\Delta} \bos{O}^{(0)} + \bos{Q}_{\bos{S}^{(1)}} \bos{O}^{(0)} + \bos{Q}_{\bos{S}^{(0)}} \bos{O}^{(1)}&=0,\cr
\vdots\quad &
}\right.
}
We  note that $\bos{Q}_{\bos{S}^{(0)}}^2=0$ due to the tree level
master equation in \bab. Thus, to find a solution $\bos{K_S}\bos{O}=0$,
we may start from   $\bos{Q}_{\bos{S}^{(0)}}\bos{O}^{(0)}=0$.
Note that $\bos{O}^\pr$ would be $\bos{K_S}$-exact if there exists
$\bos{\La}=\bos{\La}^{(0)}+\sum_{n=1}^\infty \hbar^n \bos{\La}^{(n)}$
such that
\eqn\bas{
\bos{O}^\pr=\bos{K_S}\bos{\La} \quad \Longrightarrow
\left\{\eqalign{
 \bos{Q}_{\bos{S}^{(0)}}\bos{\La}^{(0)}&=\bos{O}^{\pr(0)},\cr
-\bos{\Delta} \bos{\La}^{(0)} + \bos{Q}_{\bos{S}^{(1)}} \bos{\La}^{(0)} + \bos{Q}_{\bos{S}^{(0)}} \bos{\La}^{(1)}&=\bos{O}^{\pr(1)},\cr
\vdots\quad &
}\right.
}
Thus to find  $\bos{K_S}$ cohomology classes we may start from
cohomology of the following complex
\eqn\bata{
\left(\bos{Q}_{\bos{S}^{(0)}}, \mT =\bigoplus_k \mT_k\right).
}
But we see that building a representative $\bos{K_S}$ cohomology class
from a $\bos{Q}_{\bos{S}^{(0)}}$ cohomology class is an elaborated
as well as complicated procedure. We may ask why QFT instruct
us to do such procedure, which an answer would be given, implicitly,
in a later part of this notes.

We shall call an element of cohomology of the complex \bata\  a classical
observable, while an element of cohomology of the complex
\aam\ quantum observables. We may also call 
a solution $\bos{S}^{(0)}$ of the classical master equation
\tree\ classical (BV) action functional while a solution
$\bos{S}$ of the master equation \master\ quantum (BV) action functional.
We may say that quantizablity of a classical theory
means that (assuming the existence of $\bos{\Delta}$), the existence
of quantum (BV) action functional as well as
a {\it quasi-isomorphism} between
the quantum BV complex \aam\ to the classical
complex \bata. 

Now we shall clarify the last statement.
Consider a classical observable $\bos{O}^{(0)}$.
The $1$st obstruction $\mathfrak{B}^{(1)}$ to have the corresponding quantum observable
is, from \bana;
\eqn\qna{
\mathfrak{B}^{(1)}=-\bos{\Delta} \bos{O}^{(0)} + \bos{Q}_{\bos{S}^{(1)}} \bos{O}^{(0)} + \bos{Q}_{\bos{S}^{(0)}} \bos{O}^{(1)}.
}
We can show that $\bos{Q}_{\bos{S}^{(0)}}\mathfrak{B}^{(1)}=0$
as follows; to begin with we have $\left(\bos{S}^{(0)}, \bos{O}^{(0)}\right)=0$
implying
\eqn\qnb{
\left(\bos{\Delta}\bos{S}^{(0)}, \bos{O}^{(0)}\right)
+\left(\bos{S}^{(0)}, \bos{\Delta}\bos{O}^{(0)}\right)=0,
}
as $\bos{\Delta}$ is a derivation of the BV bracket.
{}From the quantum master equation $\bos{\Delta}\bos{S}^{(0)}=\left(\bos{S}^{(0)},\bos{S}^{(1)}\right)$,
we have
\eqn\qnc{
\left(\left(\bos{S}^{(0)},\bos{S}^{(1)}\right), \bos{O}^{(0)}\right)
+\left(\bos{S}^{(0)}, \bos{\Delta}\bos{O}^{(0)}\right)=0.
}
Now we note that
\eqn\qnd{
\bos{Q}_{\bos{S}^{(0)}}\mathfrak{B}^{(1)}\equiv
\left(\bos{S}^{(0)},\mathfrak{B}^{(1)}\right)=
-\left(\bos{S}^{(0)},\bos{\Delta} \bos{O}^{(0)}\right)
+\left(\bos{S}^{(0)}, \left(\bos{S}^{(1)}, \bos{O}^{(0)}\right)\right)
}
{}From $\left(\bos{S}^{(0)}, \bos{O}^{(0)}\right)=0$ and 
the Jacobi-identity of BV barcket, we have
\eqn\qne{
\left(\bos{S}^{(0)}, \left(\bos{S}^{(1)}, \bos{O}^{(0)}\right)\right)
= - \left(\left(\bos{S}^{(0)},\bos{S}^{(1)}\right), \bos{O}^{(0)}\right).
}
Thus we have $\bos{Q}_{\bos{S}^{(0)}}\mathfrak{B}^{(1)}=0$.
Then the situation is exactly like the $1$-loop anomaly
we discussed before, and the absense of $1$-loop anomaly
implies that $\mathfrak{B}^{(1)}=0$. The similar argument
can be extended to all the higher order terms in $\hbar$.
Consequently the existence of quantum BV action functional
also implies existence of quasi-isomorphism between
the classical and quantum BV complexes as well.

\subsection{Family of QFTs}

Now we are going to show that 
the deformed solution 
$\bos{S}^{(0)}(t)$, defined by \msrk, of the classical master equation 
can be also quantum corrected as follows
\eqn\msrmb{
\eqalign{
\bos{S}(t) =&\bos{S}^{(0)}(t) +\sum_{\ell=1}^\infty\hbar^\ell \bos{S}^{(\ell)}(t)\cr
=&\bos{S}^{(0)}_o +  t^a\bos{O}^{(0)}_a +\sum_{n=2}^\infty \Fr{1}{n!}t^{a_1}\ldots t^{a_n}
\bos{O}^{(0)}_{a_1\ldots a_n}\cr
&+\sum_{\ell=1}^\infty\hbar^\ell\left(\bos{S}^{(\ell)}_o +  t^a\bos{O}^{(\ell)}_a +\sum_{n=2}^\infty \Fr{1}{n!}t^{a_1}\ldots t^{a_n}
\bos{O}^{(\ell)}_{a_1\ldots a_n}\right)
}
}
to satisfy quantum master equation if $H^1_{\bos{Q}_{\bos{S}^{(0)}_o}}$
is trivial.

The $1$-loop anomaly for the deformed theory is
\eqn\msrmc{
\bos{\mathfrak{A}}^{(1)}(t) = \bos{\Delta}\bos{S}^{(0)}(t) -\bos{Q}_{\bos{S}^{(0)}(t)}\bos{S}^{(1)}(t),
}
where $\bos{Q}_{\bos{S}^{(0)}(t)} =\left(\bos{S}^{(0)}(t),\ldots\right)$
and
$\bos{\mathfrak{A}}^{(1)}(t) =\bos{\mathfrak{A}}^{(1)}_o +\sum_{n=1}^\infty\Fr{1}{n!}t^{a_1}\ldots
t^{a_n} \bos{\mathfrak{A}}^{(1)}_{a_1\ldots a_n}$. 
We have
\eqn\msrme{
\eqalign{
\bos{\mathfrak{A}}^{(1)}_o& = \bos{\Delta}\bos{S}^{(0)}_o -\bos{Q}_{\bos{S}^{(0)}_o}\bos{S}^{(1)}_o,\cr
t^a\bos{\mathfrak{A}}^{(1)}_a &=t^a\bos{\Delta}\bos{O}^{(0)}_a - t^a\left(\bos{O}^{(0)}_a, \bos{S}^{(1)}_o\right)
-t^a \bos{Q}_{\bos{S}^{(0)}_o}\bos{O}^{(1)}_a,\cr
t^a t^b\bos{\mathfrak{A}}^{(1)}_{ab} &=t^a t^b\bos{\Delta}\bos{O}^{(0)}_{ab} 
- t^a t^b\left(\bos{O}^{(0)}_a, \bos{O}^{(1)}_b\right)
- t^a t^b\left(\bos{O}^{(0)}_{ab}, \bos{S}^{(1)}_o\right)
-t^a t^b\bos{Q}_{\bos{S}^{(0)}_o}\bos{O}^{(1)}_{ab},\cr
}
}
etc.
Based on the relation
$\left(\bos{S}^{(0)}(t),\bos{S}^{(0)}(t)\right)=0$
and
$
\left(\bos{S}^{(0)}(t),\bos{\Delta S}^{(0)}(t)\right)=0$,
it can be shown that
\eqn\msrmf{
\eqalign{
\bos{Q}_{\bos{S}^{(0)}_o}\bos{\mathfrak{A}}^{(1)}_o &=0,\cr
\bos{Q}_{\bos{S}^{(0)}_o}\bos{\mathfrak{A}}^{(1)}_a 
&= (\bos{O}^{(0)}_a, \bos{\mathfrak{A}}_o),\cr
\bos{Q}_{\bos{S}^{(0)}_o}\bos{\mathfrak{A}}^{(1)}_{ab} &= 
\left(\bos{O}^{(0)}_{ab}, \bos{\mathfrak{A}}^{(1)}_o\right) +
\left(\bos{O}^{(0)}_{a}, \bos{\mathfrak{A}}^{(1)}_b\right) +
\left(\bos{\mathfrak{A}}^{(1)}_{a}, \bos{O}^{(0)}_{b}\right), 
}
}
etc.
Since
$H^1_{\bos{Q}_{\bos{S}^{(0)}_o}}=0$
leading to $\bos{\mathfrak{A}}^{(1)}_o =0$, we have 
$\bos{Q}_{\bos{S}^{(0)}_o}\bos{\mathfrak{A}}^{(1)}_a=0$
leading to $\bos{\mathfrak{A}}^{(1)}_a =0$. 
Using $\bos{\mathfrak{A}}^{(1)}_o =\bos{\mathfrak{A}}^{(1)}_a=0$ we have
$\bos{Q}_{\bos{S}^{(0)}_o}\bos{\mathfrak{A}}^{(1)}_{ab}=0$
leading to $\bos{\mathfrak{A}}^{(1)}_{ab} =0$. Using induction 
it can be shown that $\bos{\mathfrak{A}}^{(1)}(t) =0$.
Adopting the similar procedure it can be shown that
the all-loop anomaly $\bos{\mathfrak{A}}(t) =\sum_{\ell=1}^\infty
\bos{\mathfrak{A}}^{(\ell)}(t)$ also vanishes.

In summary we see that two possible obstruction of the
moduli space $\mathfrak{N}$ in the neighborhood of $o \in \mathfrak{N}$
are  $H^{-1}_{\bos{Q}_{\bos{S}^{(0)}_o}}$ and
$H^1_{\bos{Q}_{\bos{S}^{(0)}_o}}$, which are related with
symmetry and anomaly, respectively.
Assuming the small resolution the
moduli space $\mathfrak{N}$ in the neighborhood of $o \in \mathfrak{N}$
is unobstructed and the tangent space $T_o\mathfrak{N}$ is
isomorphic to $H^0_{\bos{Q}_{\bos{S}^{(0)}_o}}$ provided
that there is no anomaly. In other words the absence of anomaly
in QFT implies that a QFT come with  family, which property shall be
crucial in the next section.

\subsubsection{Generalization}

Now we consider an natural generalization the above picture. 
We may
decompose the space $\mT=\mT_{even}\oplus \mT_{odd}$ into the $U=even$
and $U=odd$ subspaces. We have
\eqn\msrn{
\left(\star,\star\right):\mT_{odd}\otimes\mT_{odd}\longrightarrow \mT_{odd},
}
which means that the BV bracket endows a structure of graded Lie algebra
on $\mT_{odd}$.  Then we have the parity preserving adjoint action 
generated by elements of $\mT_{odd}$. Now we allow classical BV action functional
to be an element of $\mT_{even}$ in general and
define extended moduli space $\mathfrak{M}$ by
\eqn\msrh{
\mathfrak{M} =\left\{\bos{S}^{(0)}\in \mT_{even}| \left( \bos{S}^{(0)},
\bos{S}^{(0)}\right)=0/\sim\right\}
} 
where the equivalence $\sim$ is defined by the adjoint action by elements
in $\mT_{odd}$. 

We choose a base point $o \in \mathfrak{N}\subset \mathfrak{M}$
and the corresponding classical BV action functional
$\bos{S}^{(0)}_o \in \mT_0 \subset \mT_{even}$.
Let $\{\bos{\Upsilon}^{(0)}_m\}$ be a basis of $H^{odd}_{\bos{Q}_{\bos{S}^{(0)}_o}}$.
Then we have the structure of Lie algebra on 
$H^{odd}_{\bos{Q}_{\bos{S}^{(0)}_o}}$
such that
\eqn\msro{
\left(\bos{\Upsilon}^{(0)}_m,\bos{\Upsilon}^{(0)}_n\right) = f^r_{mn}\bos{\Upsilon}^{(0)}_r
\hbox{ mod } \Im  \bos{Q}_{\bos{S}^{(0)}}
,
}
with certain structure "constants" $f^r_{mn}=-f^r_{nm}$..\foot{Each component
of $\{f_{mn}^r\}$ can be even function, in general, with ghost number $U=even$.
The Jacobi identity of BV bracket suggest that
one may regard the pairts $(\{f_{mn}^r\},\{(\bos{\Upsilon}^{(0)}_m\})$
as structure of graded Lie algebroid over graded space.}
We extend $\bos{\CT}$ to
\eqn\msrp{
\bos{\tilde\CT}=\bos{\CT}\times T^*[-1]\left( \prod_{k\in \Z}
H^{-(2k+1)}_{\bos{Q}_{\bos{S}^{(0)}_o}}[2k+1]\right)
}
by introducing generalized ghosts {\tt fields} $\{C^m\}$
such that
\eqn\msrq{
U(C^m) +U(\bos{\Upsilon}^{(0)}_m) =0,\qquad
|C^m| +|\bos{\Upsilon}^{(0)}_m| =0,
}
and their {\tt antifields} $\{C^\bullet_m\}$.
Then we extend $\bos{S}^{(0)}_o$ to $\bos{\tilde S}^{(0)}_o$;
\eqn\msrs{
\bos{\tilde S}^{(0)}_o = \bos{S}^{(0)}_o + C^m\bos{\Upsilon}^{(0)}_m +\Fr{1}{2}C^m C^n f^r_{mn}C^\bullet_r
}
which satisfies the classical master equation. 
Then we need to check if 
$H^{odd}_{\bos{\tilde Q}_{\bos{\tilde S}^{(0)}_o}}$ is trivial. Otherwise
we need to repeat the above procedure until we get trivial
cohomology for odd elements.

We may call the above procedure  an extended resolution.
{}From now on we assume that  $(\bos{\CT},\bos{S}^{(0)})$
had been already resolved in the extended sense.
Let $\{\bos{O}^{(0)}_\a\}$ be a basis of 
$\oplus H^{\bullet}_{\bos{Q}_{\bos{S}^{(0)}_o}}=H^{even}_{\bos{Q}_{\bos{S}^{(0)}_o}}$
and  $\{t^a\}$ be the dual basis with 
$U(t^\a) +U(\bos(O)^{(0)}_\a)=0$ and
$|t^\a|+|\bos{O}^{(0)}_\a|=0$. Then the tangent space
$T_o\mathfrak{M}$ is isomorphic to $\oplus_\bullet  H^{\bullet}_{\bos{Q}_{\bos{S}^{(0)}_o}}$
and there exist another solution of classical master equation given by
\eqn\msrt{
\bos{\CS}^{(0)}=\bos{S}^{(0)}_o +  t^\a\bos{O}^{(0)}_\a +\sum_{n=2}^\infty \Fr{1}{n!}t^{\a_1}\ldots t^{\a_n}
\bos{O}^{(0)}_{\a_1\ldots \a_n},
}
which can be found by following the similar procedure described before.
Applying the same reasoning as before we conclude that there exist
solution $\bos{\CS}= \bos{\CS}^{(0)} +\hbar \bos{\CS}^{(1)} +\CO(\hbar^2)$
of quantum master equation.
Thus the extended moduli space $\mathfrak{M}$ is smooth in the neighborhood
of $o$.

\subsection{The BV Quantization II}

End of digression and we turn to the partition function
of the theory is defined by
\eqn\bal{
\bos{Z}(\hbar)_{[\bos{\CL}]} = \int_{\bos{\CL}}d\m\; e^{-\bos{S}/\hbar}, 
}
where $\bos{\CL}$ is a Lagrangian subspace and $d\m$ 
is its "measure". The master equation is the condition that
the $Z(\hbar)$ depends only on the homology class 
$[\bos{\CL}]$ of
$\bos{\CL}$.\foot{ On might argue that the "measure" $d\m$ is
ambiguous and ill defined. We should, however, note that
$\bos{\CL}$, the space of all {\tt fields} that we start with, is typically affine (linear) graded space, which
has "standard measure", $\prod d\phi^A$, though infinite product. 
Furthermore the actual measure is $d\m e^{-\bos{S}/\hbar}\bigl|_{\bos{\CL}}$,
which may be viewed as a top form on $\bos{\CL}$. In other
words the BV quantization scheme want to achieve 
good notion of cohomology class by carefully defining $\bos{S}$.}
The master equation for $\bos{S}$ also implies that
the partition function is independent to {\it canonical transformation}
connected to identity generated by so called 
{\it gauge fermion}; 
\eqn\balb{
\bos{\G}=\bos{\G}^{(0)} +\sum_{\ell=1}^\infty \hbar^\ell \bos{\G}^{(\ell)}\in \mL[[\hbar]]_{-1},
}
which is an odd function, with ghost number $U=-1$,
on $\bos{\CL}$ in formal power
series in $\hbar$. Then we may choose suitable representative
in  the middle dimensional cohomology class $\left[d\m\; e^{-\bos{S}/\hbar}\right]$
in $\bos{\CT}$ such that the path integral has good behavior.
Equivalently one can choose suitable representative in the
middle dimension homology class $[\bos{\CL}]$.
The above procedure is called gauge fixing.
We note that the two inequivalent QFTs may share the
same or equivalent BV action functional and $\bos{\CT}$, while
correspond to choosing two inequivalent Lagrangian subspaces.
We note that the problem we are dealing with has
close analogue in defining pairings between
homology and cohomology classes in the finite dimensional manifold.

Thus, in practice, we may choose a simple Lagrangian subspace
$\bos{\CL}$, like the space of all initial {\tt fields}, and a good 
gauge fermion $\bos{\G}\in \mL[[\hbar]]_{-1}$. After the canonical
transformation we have so called gauge fixed action functional
\eqn\bala{
\eqalign{
S_{[\bos{\CL}]} 
=& \bos{m}_0 + \bos{m}_1(\bos{\G}) 
+\sum_{n=2}^\infty\Fr{1}{n!} \bos{m}_n\left(\bos{\G},\cdots,\bos{\G}\right)
\cr
=&
\bos{s} + \bos{q}\bos{\G}^{(0)}  
+\sum_{n=3}^\infty\Fr{1}{n!} \bos{m}^{(0)}_n\left(\bos{\G}^{(0)},\cdots,\bos{\G}^{(0)}\right)\cr
&
+\CO(\hbar),
}
} 
where $S_{[\bos{\CL}]} :=\bos{S^\G}|_{\bos{\CL}}\in \mL[[\hbar]]_0$, so that
$Z(\hbar)_{[\bos{\CL}]} = \int_{\bos{\CL}}d\m\; e^{-{S}_{[\bos{\CL}]}/\hbar}$.

Remark that in the semi-classical limit $\hbar\rightarrow 0$ and
the case that $\bos{m}^{(0)}_{n}=0$ for $\forall n\geq 2$ the gauge fixed action functional
becomes $\bos{s} +\bos{q\G}^{(0)}$ exactly as in the familiar BRST quantization.
In such a case we have $\bos{qs}=\bos{q}^2=0$ by construction.
The above also show that the path integral is independent to gauge
choice in the BRST quantization, which proof seems to
be the original motivation of Batalin-Vilkovisky.

We also note that it may also possible that the gauge fixed action
functional \bala\ can be zero;
\eqn\balaz{
 \bos{m}_0 + \bos{m}_1(\bos{\G}) 
+\sum_{n=2}^\infty\Fr{1}{n!} \bos{m}_n\left(\bos{\G},\cdots,\bos{\G}\right)
=0,
}
i.e., the Maurer-Cartan equation of quantum weakly homotopy Lie $(-1)$-algebroid.
The above condition implies that
the operator $(\bos{m}_1)_{\bos{\G}}$ defined by
\eqn\balay{
(\bos{m}_1)_{\bos{\G}}:= \bos{m}_1 
+\sum_{n=2}^\infty\Fr{1}{(n-1)!} \bos{m}_n\left(\bos{\G},\cdots,\bos{\G},\phantom{\G}\right)
}
satisfies that 
\eqn\balyx{
(\bos{m}_1)^2_{\bos{\G}}=0.
}
Then $(\bos{m}_1)_{\bos{\G}}$ may be used to define physical states
by its cohomology.\foot{In Zwiebach's string field theory \cite{Z} 
non-zero $\bos{s}=\bos{m}_0^{(0)}$
in $\bos{m}_0 =\sum_{\ell=0}^\infty \hbar^\ell \bos{m}_0^{(\ell)}$
corresponds to non-conformal background. Then the condition
\balaz\ may be interpreted as the equation for moduli space of
conformal background. We also remark that an equation similar
to the semi-classical version of the equation \balaz\ also appears in the Floer homology
of Lagrangian intersection in the work of Fukaya et.\ als. \cite{FOOO} as condition for vanishing obstruction for Floer homology, where the semi-classical version of 
$(\bos{m}_1)_{\bos{\G}}$ is interpreted as correct Floer boundary operator.
We also remark that the semi-classical in the above contexts correpond
to genus zero in string theory.
}

The similar argument can be applied to expectation value of observable.
An observable is $\bos{K_S} =-\hbar\bos{\Delta} + \bos{Q_S}$ closed function
$\bos{O}=\bos{O}^{(0)}+\sum_{n=1}^\infty \hbar^n \bos{O}^{(n)}\in\mT[[\hbar]]$ on $\bos{\CT}$ in the formal power series of $\hbar$;
\eqn\ban{
\bos{K_S}\bos{O} =0.
}
The {\it expectation value} of such an $\bos{O}$ is denoted by
\eqn\baq{
\left<\bos{O}\right>_{[\bos{\CL}]} = \int_{\bos{\CL}}d\m\; \bos{O} e^{-\bos{S}/\hbar}
}
Then we have a fundamental identity
\eqn\bar{
\left<\bos{O}+ \bos{K_S}\bos{\Lambda}\right>_{[\bos{\CL}]} 
=\left<\bos{O}\right>_{[\bos{\CL}]} 
}
namely the expectation value depends only on the $\bos{K_S}$
cohomology class (and on the homology class of $[\bos{\CL}]$ of $\bos{\CL}$. 
The above identity of BV is 
generalization of identities in QFT
like those of Slavanov-Taylor, Ward-Takasaki and Zinjustin.
We may take an alternative view such that the above identity 
as a necessary condition to have a proper definition
of path integral, namely a representative of homology class 
$[\bos{\CL}]$ should be chosen to ensure 
$\left<\bos{K_S}\bos{\Lambda}\right>_{[\bos{\CL}]} =0$.
It is clear that two inequivalent QFTs may have the
equivalent BV model but correspond to two inequivalent Lagrangian
subspaces as the space of {\tt fields}. More precisely path integrals
of a BV model
define certain "periods"\foot{It may be worth to remark that
path integrals of QFT with finite dimensional field space (or $0+0$
dimensional QFT) is closely related to the exponential period considered
briefly in the exposition of Kontsevich-Zagier \cite{KZ}. The quantum flat
structure we will talk about in the next section, then, may be translated into
theory of (extended) variations of exponential periods.}
matrix of pairing between homology and
cohomology classes, while each of its matrix element has been,
traditionally, referred as path integral of a certain QFT.

\newsec{Reaching Out To One's Family}

The BV quantization scheme suggests us to
work out all $\bos{K_S}$ cohomology classes (quantum observables) and
study their expectation values via path integral.
One of the most crucial purpose of QFT is to
understand correlation functions, which are, naively, 
the expectation values of products of observables.  
Here, the BV quantization scheme introduce
a vexing problem that products of quantum observables are not quantum 
observables in general (comments below \aam\ that products
of $\bos{K_S}$ cohomology classes are not, in general,
Kernel of $\bos{K_S}$\foot{This also implies that
one should work at the level of quantum BV complex
rather than its cohomology, which view shall not be
elaborated in this notes.}
), meaning that
the expectation value of products of quantum observables may depend on
continuous variations of Lagrangian subspace (or gauge choice).
We shall see the above problem is actually a clue suggesting that
we need  to find  family QFTs.
The upshot is that
QFTs come with a family parametrized by certain
moduli space and every path integrals of a given QFT belonging
to such a family means reaching out to
its neighborhood. If so 
QFT(s) should eventually be understood
in its totality.

Let $\{\bos{O}_\a\}$ be a basis of all $\bos{K_S}$ cohomology classes among
elements in $\mT[[\hbar]]$. 
Being an observable, $\bos{K_S}\bos{O}_\a =0$,
we have
\eqn\bba{
\bos{\Delta}\left(\bos{O}_\a e^{-\bos{S}/\hbar}\right) =0, \quad\hbox{since}\quad
\bos{\Delta} e^{-\bos{S}/\hbar} =0.
}
The above relation strongly suggests that evaluation the
expectation values of observables corresponds to infinitesimal
deformation of given QFT, characterized by given $\bos{S}$ and
$\bos{\CT}$. Let's consider the following would be "generating functional"
\eqn\bbb{
Z(\hbar)_{\vec{t}} = \int_{\bos{\CL}}d\m e^{- (\bos{S} + t^\a \bos{O}_\a)/\hbar},
}
where $\{t^\a\}$ be the dual basis of $\{\bos{O}_\a\}$ such
that $|t^\a| = -|\bos{O}_\a|$.
Then the expectation value of observable is
\eqn\bbc{
\left<\bos{O}_\a\right> = -\hbar\Fr{\rd Z_t}{\rd t^\a}\biggl|_{\{\vec{t}\}=\{\vec{0}\}}
}
Thus the "generating functional" might contains all the information of
expectation values. Now we consider the expectation
value of products of two observables, which may be written as
\eqn\bbd{
\left<\bos{O}_\a\bos{O}_\b\right> =\hbar^2\Fr{\rd^2 Z_t}{\rd t^\b\rd t^\a}\biggl|_{\{\vec{t}\}=\{\vec{0}\}}.
}
We recall, however, that  the product  of two observables  is not an observable in
general as $\bos{K_S}$ is not a derivation of product, which obstruction
is measured by the BV bracket $\left(\bos{O}_\a,\bos{O}_\b\right)$.
Thus the naive definition \bbb\ of generating functional is {\it not correct}.
The problem is that the path integral  \bbb\ is ill defined
as the $\bos{S} + t^\a\bos{O}_\a$ satisfies the master equation, in general, only up to
the first order in $\{t^\a\}$.

Instead we seek for solution $\bos{S}(\{t^\a\})$ of the master
equation
\eqn\bbe{
-\hbar \bos{\Delta} \bos{S}(\{t^a\}) + \Fr{1}{2}\biggl(\bos{S}(\{t^a\}),\bos{S}(\{t^\a\})\biggr)=0.
}
such that
\eqn\bbf{
 \bos{S}(\{t^a\}) =\bos{S} +t^\a\bos{O}_\a + \sum_{n=2}^\infty\Fr{1}{n!} t^{\a_1}\cdots t^{\a_n}
 \bos{O}_{\a_1\cdots\a_n},\qquad
|\bos{S}(\{t^a\})|=0,
}
where $\bos{S}$ may be regarded the initial condition and $\{\bos{O}_\a\}$ as infinitesimal
deformations. 
Then we have an well-defined generating functional
\eqn\bbg{
\bos{Z}(\{t^\a\}) = \int_{\bos{\CL}}d\m\; e^{- \bos{S}(\{t^\a\})/\hbar},
}
which contains all information of QFT defined by the BV action functional
$\bos{S} =\bos{S}(\vec{t}=\vec{0})$.\foot{
At this point one might ask if solutions of master equation \bbe\ 
always exists for a given field theory such that the first order is spanned by a basis
of all $\bos{K_S}$-cohomology classes. The answer would be certainly
"No" in general, but the answer should be "Yes" for a QFT that, if otherwise,
our starting point is wrong or we need some modifications and extensions
until the answer is Yes (see Section $2.4$). This is part of our demand of 
"renormalized" quantizablity.} 

For simplicity in notation we shall denote $\bos{\CS}=\bos{S}(\{t^\a\})$
and $\bos{\CZ} =\bos{Z}(\{t^\a\})$, such that
\eqn\bbh{
\bos{\CZ} =  \int_{\bos{\CL}}d\m\; e^{- \bos{\CS}/\hbar},\qquad
-\hbar \bos{\Delta} \bos{\CS} + \Fr{1}{2}(\bos{\CS},\bos{\CS})=0,\qquad \bos{\CS}|_{\vec{t}=\vec{0}}=\bos{S}.
}
Note that the master equation implies
that 
\eqn\bbha{
\bos{\CK_{\CS}} := -\hbar\bos{ \Delta} + \bos{\CQ_{\CS}},
\quad\hbox{where } \bos{\CQ_{\CS}}:=\left(\bos{\CS},\ldots\right)
}
satisfies 
\eqn\bbhb{
\bos{\CK}_{\bos{\CS}}^2=0.
}
Then  we have the corresponding  identity;
\eqn\bbr{
\int_{\bos{\CL}}d\m\;\left(\bos{\CK_\CS}\hbox{\it (anything)}\right)e^{-\bos{\CS}/\hbar}
=0.
}
Now let's apply  derivative $\Fr{\rd}{\rd t^\a}$ to the master equation
(the middle equation in \bbh)
to obtain
\eqn\bbi{
-\hbar\bos{\Delta} \bos{\CS}_\a + \left(\bos{\CS}, \bos{\CS}_\a\right)=0
}
where $\bos{\CS}_\a := \Fr{\rd \bos{\CS}}{\rd t^\a}$.
Thus $\{\bos{\CS}_\a\}$ span, at least, subspace of $\bos{\CK_\CS}$ cohomology group.
It is obvious that $
\bos{\CK_\CS\CS}_\a\biggl|_{\vec{t}=\vec{0}} =0$,
leading to  $\bos{K_S O}_\a =0$,
where $\bos{O}_\a =\bos{\CS}_\a|_{\vec{t}=\vec{0}}$.

{}From the definition of $\bos{\CZ}$ we have
\eqn\bbk{
\hbar \bos{\CZ}_\a = -\int_{\bos{\CL}} d\m\; \bos{\CS}_\a e^{-\bos{\CS}/\hbar}
}
where
\eqn\bbl{
\bos{\CZ}_\a := \Fr{\rd \bos{\CZ}}{\rd t^\a}.
}

{}From \bbk\ we also have
\eqn\bbm{
\hbar^2 \Fr{\rd \bos{\CZ}_\a}{\rd t^\b} 
=  \int_{\bos{\CL}} d\m\ 
\left( -\hbar \Fr{\rd\bos{\CS}_\a}{\rd t^\b} + \bos{\CS}_\b\bos{\CS}_\a\right)
e^{-\bos{\CS}/\hbar}
}
Then we check if the expression 
$\left(
-\hbar \Fr{\rd\bos{\CS}_\a}{\rd t^\b} + \bos{\CS}_\b\bos{\CS}_\a\right)$ 
is also $\bos{\CK_\CS}$-closed. 
{}From \bbi, after taking derivative $\Fr{\rd}{\rd t^\b}$ we  have
\eqn\bbn{
-\hbar\bos{\Delta} \Fr{\rd \bos{\CS}_\a}{\rd t^\b} + \left(\bos{\CS},\Fr{\rd \bos{\CS}_\a}{\rd t^\b}\right)
+\left(\bos{\CS}_\b,\bos{\CS}_\a\right)=0.
}
It follows, further using \aaf\ and \bbi, that 
\eqn\bbna{
\bos{\CK_\CS}\left(
-\hbar \Fr{\rd\bos{\CS}_\a}{\rd t^\b} + \bos{\CS}_\b\bos{\CS}_\a\right)=0.
}

\subsec{Quantum Algebra of Observables at Base Point}

We may regard the master equation \bbe\ defines formal neighborhood
of the initial QFT defined by $\bos{S}$ in certain moduli space
$\bos{\mathfrak{M}}$ of family of QFTs such that the initial QFT is a
(classical) base point 
$o$ and the given basis $\{\bos{O}_a\}$ of $\bos{K_S}$-cohomology 
is a basis of tangent space $T_o\bos{\mathfrak{M}}$ at $o \in \bos{\mathfrak{M}}$.
Then the dual basis $\{t^\a\}$ of $\bos{K_S}$-cohomology  can be
regarded as a local coordinate system around the base point $o$.
Now we want to characterize expectation values of all observables
and their correlation functions of the initial QFT.

The condition \bbna\ implies that
\eqn\bpa{
\left(\bos{\CK_\CS}\left(
-\hbar \Fr{\rd\bos{\CS}_\a}{\rd t^\b} + \bos{\CS}_\b\bos{\CS}_\a\right)\right)\biggl|_{\vec{t}=0}=0.
\quad \Longrightarrow \bos{K_S}\left(
-\hbar \bos{O}_{\a\b} +\bos{O}_\b\bos{O}_\a\right)=0.
}
That is $(-\hbar \bos{O}_{\a\b} +\bos{O}_\b\bos{O}_\a)$ is
$\bos{K_S}$-closed while $(\bos{O}_\b\bos{O}_\a)$ may not.
Since $\{\bos{O}_\a\}$ is a (complete) basis of $\bos{K_S}$-cohomology
we have
\eqn\bpb{
-\hbar \bos{O}_{\a\b} +\bos{O}_\b\bos{O}_\a = \bos{A}^\g_{\a\b}\bos{O}_\g 
+ \bos{K_S}\bos{\La}_{\a\b}
}
for some structure constants $\{\bos{A}^\g_{\a\b}\}$, in formal power series
of $\hbar$, and for some $\bos{\La}_{\a\b} \in
\mathfrak{T}[[\hbar]]$.
It follows that
\eqn\bpc{
\left<\bos{O}_\b\bos{O}_\a\right>_{\vec{t}=0} 
= \bos{A}^\g_{\a\b}\left<\bos{O}_\g\right>_{\vec{t}=0} +\hbar\left<\bos{O}_{\a\b}\right>_{\vec{t}=0}.
}
We note that the structure constants $\{\bos{A}^\g_{\a\b}\}$
are independent to any choice of Lagrangian subspace.
The expectation value $\left<\bos{O}_\g\right>_{\vec{t}=0}$
depends on homology class of Lagrangian subspace,
where we integrated over. One the other hand
both  $\left<\bos{O}_\b\bos{O}_\a\right>_{\vec{t}=0}$
and $\hbar\left<\bos{O}_{\a\b}\right>_{\vec{t}=0}$ depends on
variation of Lagrangian subspace even within the same homology
class, which dependence cancel with each others.
The formula \bpc\ is very suggestive as the correlation
function (LHS of \bpc) of two observables $\bos{O}_\a$ and $\bos{O}_\b$
involve expectation values of all observables and something shared only by
two of them.\foot{Compare with the  classical limit $\hbar=0$;
$\left<\bos{O}^{(0)}_\b\bos{O}^{(0)}_\a\right> 
= \bos{A}^{(0)\g}_{\a\b}\left<\bos{O}^{(0)}_\g\right>
$, where $\bos{O}= \bos{O}^{(0)} + \hbar \bos{O}^{(1)}+\ldots$ and
$\bos{A}= \bos{A}^{(0)} + \hbar \bos{A}^{(1)}+\ldots$.}

In general, a solution \bbf\ of the master equation \bbe\
leads to "quantum" products of observables
defined by
\eqn\bpe{
\bos{a}_n\left(\bos{O}_{\a_1},\ldots,\bos{O}_{\a_n}\right) 
=e^{\bos{\CS}/\hbar}\left(\Fr{\hbar^n \rd^n}{\rd t^{\a_n}\ldots\rd t^{\a_1}}\right)
e^{-\bos{\CS}/\hbar} \biggl|_{\vec{t}=0}
}
such that $\bos{K_S} \bos{a}_n(\bos{O}_{\a_1},\ldots,\bos{O}_{\a_n})=0$,
which implies that
\eqn\bpf{
\bos{a}_n\left(\bos{O}_{\a_1},\ldots,\bos{O}_{\a_n}\right) =\bos{A}_{\a_1\ldots\a_n}^\g\bos{O}_\g
+\bos{K_S}\bos{\La}_{\a_1\ldots\a_n}.
}
For examples
we have
\eqn\bph{
\eqalign{
\bos{a}_2(\bos{O}_\a,\bos{O}_\b) =&\bos{O}_\a\bos{O}_\b-\hbar \bos{O}_{\b\a}
,\cr
\bos{a}_3\left(\bos{O}_\a,\bos{O}_\b,\bos{O}_\g\right)
=&\bos{O}_\a\bos{O}_\b\bos{O}_\g + \hbar(\bos{O}_{\a\g}\bos{O}_\b
+\bos{O}_\a\bos{O}_{\b\g} +\bos{O}_{\g\b}\bos{O}_\a)
\cr
&
-\hbar^2\bos{O}_{\g\b\a}
}
}
We note that all the structure constants $\{\bos{A}_{\a_1\ldots \a_n}\}$,
$n\geq 2$, are independent to any choice of Lagrangian subspace.

It follows that
we have a sequence of multiplication maps
from $\bos{K_S}$-cohomology to $\Ker \bos{K_S}$.
Let $\bos{H_{K_S}}$ be space of $\bos{K_S}$-cohomology classes.
Then we have the following multi-linear maps, $n\geq 2$
\eqn\bpi{
\bos{a}_n: \bos{H}_{\bos{K_S}}^{\otimes n}\longrightarrow \Ker \bos{K_S}
}
and associated $[\bos{a}_n]_{\vec{t}=0}$
\eqn\bpj{
\eqalign{
[\bos{a}_n]_{\vec{t}=0}: \bos{H}_{\bos{K_S}}^{\otimes n}&\longrightarrow \bos{H_{K_S}},
\cr
\left[\bos{a}_n\left(\bos{O}_{\a_1},\ldots,\bos{O}_{\a_n}\right)\right]_{\vec{t}=0} &=\bos{A}_{\a_1\ldots\a_n}^\g[\bos{O}_\g]_{\vec{t}=0}.
}
}
It follows that
\eqn\bpk{
\biggl< \bos{a}_n\left(\bos{O}_{\a_1},\ldots,\bos{O}_{\a_n}\right)\biggr>_{\vec{t}=0}
=\bos{A}_{\a_1\ldots\a_n}^\g\biggl<\bos{O}_\g\biggr>_{\vec{t}=0},
}
Symbolically
\eqn\bpl{
<\bos{a}_n>_{\vec{t}=0} : \bos{H}_{\bos{K_S}}^{\otimes n}\longrightarrow \Bbbk[[\hbar]]
}
which map depends on homology class of Lagrangian subspace that we
integrate over.

Consequently the path integrals can be determined completely 
by the expectation values $\{\left<\bos{O}_\a\right>_{\vec{t}=0}\}$ and all the structure constants $\{\bos{A}^\g_{\a_1\ldots \a_n}\}$,
$n\geq 2$. The former data depends  only homology class of Lagrangian
subspace, while the later data can be determined completely by working
out algebra of $\bos{K_S}$-cohomology classes without doing path integrals.
As we mentioned before inequivalent choices of Lagrangian subspaces 
as the spaces of {\tt fields} correspond to different QFTs in the traditional sense.

In this subsection we used a deformed solution master equation
with infinitesimals given by  $\bos{K_S}$-cohomology classes
rather passively. We shall see that one can do much better than that.

Now we are interested in the path integrals of the theory defined by the deformed
BV action functional $\bos{\CS}$ in \bbf. It is obvious that 
the deformed partition function $\bos{\CZ}$ in \bbg\ contains
all the information about the initial QFT. 

\subsec{Quantum Flat Structures}

Now we assume that $\{\bos{\CS}_\a\}$ form a {\it complete} basis
of $\bos{\CK_\CS}$-cohomology group, which condition, in general,
may not be true.
Then, from \bbna, there should be structure functions $\bos{A}(\{t\})$ in formal
power series of $\hbar$ 
such that
\eqn\bbp{
-\hbar \Fr{\rd\bos{\CS}_\a}{\rd t^\b} + \bos{\CS}_\b\bos{\CS}_\a 
= \bos{A}(\{t\})^\g_{\b\a}\bos{\CS}_\g
+\bos{\CK_\CS}{\bos{\La}}(\{t\})_{\b\a}
}
for some ${\bos{\La}}(\{t\})_{\b\a}$
For simplicity in notation we shall denote $\bos{\CA}=  \bos{A}(\{t^\a\})$
(note that $\bos{\CA}|_{\vec{t}=\vec{0}}=\bos{A}$, which appeared
in \bpb.)

{}Using the identity \bbr\ and from \bbk, \bbm\ and \bbp\ we arrive at a fundamental
equation
\eqn\bbq{
 \left(\Fr{\rd^2}{\rd t^\a\rd t^\b} +\Fr{1}{\hbar}\bos{\CA}_{\a\b}^\g\Fr{\rd}{\rd t^\g}\right)
 \bos{\CZ}=0, \quad \Longrightarrow \quad
 \Fr{\rd \bos{\CZ}_\b}{\rd t^\a} +\Fr{1}{\hbar}\bos{\CA}_{\a\b}^\g\bos{\CZ}_\g =0.
 }
We should emphasis that the equations \bbq\  valid universally
for any quantum field theory under our present assumption.

Note that the following relation is obvious by definition;
\eqn\bbss{
\bos{\CA}_{\a\b}^\g = \bos{\CA}_{\b\a}^\g.
}
Now starting from \bbq\ we have
\eqn\bbsa{
\eqalign{
\Fr{\rd^3\bos{\CZ}}{\rd t^\g\rd t^\a\rd t^\b} 
&= -\Fr{1}{\hbar}\left(\Fr{\rd\bos{\CA}_{\a\b}^\s}{\rd t^\g}\right) \bos{\CZ}_\s
-\Fr{1}{\hbar} \bos{\CA}_{\a\b}^\r \Fr{\rd\bos{\CZ}_\r}{\rd t^\g}
\cr
 &= -\Fr{1}{\hbar}\left(\Fr{\rd\bos{\CA}_{\a\b}^\s}{\rd t^\g}\right) \bos{\CZ}_\s
+\Fr{1}{\hbar^2}\left( \bos{\CA}_{\a\b}^\r \bos{\CA}_{\r\g}^\s\right) 
\bos{\CZ}_\s.
}
}
and the similar manipulation gives
\eqn\bbsb{
\eqalign{
\Fr{\rd^3\bos{\CZ}}{\rd t^\a\rd t^\g \rd t^\b}
&=
-\Fr{1}{\hbar}\left(\Fr{\rd\bos{\CA}_{\g\b}^\s}{\rd t^\a}\right) \bos{\CZ}_\s
+\Fr{1}{\hbar^2}\left(\bos{\CA}_{\g\b}^\r \bos{\CA}_{\r\a}^\s\right) 
\bos{\CZ}_\s.
}
}
On the other hand we have the following obvious identity
\eqn\bbt{
\eqalign{
\Fr{\rd^3\bos{\CZ}}{\rd t^\a\rd t^\g \rd t^\b}
&=\Fr{\rd^3\bos{\CZ}}{\rd t^\g\rd t^\a\rd t^\b}.
}
}
Thus we obtain another universal equation
\eqn\bbu{
\eqalign{
\Fr{\rd\bos{\CA}_{\g\b}^\s}{\rd t^\a}- \Fr{\rd\bos{\CA}_{\a\b}^\s}{\rd t^\g} 
+\Fr{1}{\hbar}\left(
\bos{\CA}_{\a\b}^\r \bos{\CA}_{\r\g}^\s -\bos{\CA}_{\b\g}^\r
 \bos{\CA}_{\r\a}^\s
\right)
=0.
}
}

Without loss of generality we may assume that
there is an identity among $\{\bos{O}_\a\}$,
say $\bos{O}_0 = 1$, as $\bos{\Delta}1=\bos{Q}1=0$ and the
identity can not be exact. It follows that $\bos{\CS}_0 =\bos{O}_0$
leading to
\eqn\bbra{
\bos{\CZ}_0:=\Fr{\rd\bos{\CZ}}{\rd t^0}=-\Fr{1}{\hbar}\bos{\CZ},
}
Now the relation \bbra\ implies another relation
\eqn\bbrb{
\Fr{\rd \bos{\CZ}_\a}{\rd t^0} = -\Fr{1}{\hbar}\bos{\CZ}_\a.
}
Using \bbq, the equation \bbrb\  implies that
\eqn\bbrd{
\bos{\CZ}_\a =\bos{\CA}_{0\a}^\s \bos{\CZ}_\s
\quad\Longrightarrow\quad \Fr{\rd \bos{\CZ}_\a}{\rd t^\b} 
= \Fr{\rd}{\rd t^\b}\left(\bos{\CA}_{0\a}^\s \bos{\CZ}_\s\right).
}
Using \bbq\ one more time we have another identity
\eqn\bbqa{
\bos{\CA}_{\a\b}^\s = 
\bos{\CA}_{0\a}^\r\bos{\CA}_{\r\b}^\s
-\hbar\Fr{\rd \bos{\CA}_{0\a}^\s}{\rd t^\b} 
}
{}From the second equation of \bbrd, and from \bbrb,
we have $\Fr{\rd \bos{\CZ}_\a}{\rd t^0}=-\Fr{1}{\hbar}\bos{\CZ}_\a
=\Fr{\rd}{\rd t^0}\left(\bos{\CA}_{0\a}^\s \bos{\CZ}_\s\right)$,
which implies that
\eqn\bbqc{
\bos{\CZ}_\a =\left(
\bos{\CA}_{0\a}^\r\bos{\CA}_{\r 0}^\s
-\hbar\Fr{\rd \bos{\CA}_{0\a}^\s}{\rd t^0}\right)\bos{\CZ}_\s
}

We shall call above structure {\it Quantum Flat Structure}.

For a QFT which formal neighborhoods has quantum flat structure
we may, if one wants to, forget about path integral
representation of $\bos{\CZ}$ and work with the system of differential
equations \bbq. The set of all independent solutions of the system
may be interpreted as path integrals of inequivalent choice
of Lagrangian subspaces.

We note that 
\eqn\xxe{
\bos{\ma}_n\left(\bos{\CS}_{\a_1},\ldots,\bos{\CS}_{\a_n}\right) 
:=e^{\bos{\CS}/\hbar}\left(\Fr{\hbar^n \rd^n}{\rd t^{\a_1}\ldots\rd t^{\a_n}}\right)
e^{-\bos{\CS}/\hbar}
}
define multi-linear maps
\eqn\xxf{
\eqalign{
\bos{\ma}_n:\bos{H}^{\otimes n}_{\bos{\CK_\CS}}\longrightarrow \Ker \bos{\CK_\CS}.
}
}
Explicitly
\eqn\xxg{
\eqalign{
\bos{\ma}_2(\bos{\CS}_\a,\bos{\CS}_\b) =& -\hbar\Fr{\rd \bos{\CS}_\b}{\rd t^\a}
+\bos{\CS}_\a\bos{\CS}_\b,\cr
\bos{\ma}_3(\bos{\CS}_\a,\bos{\CS}_\b,\bos{\CS}_\g) =&
 -\hbar^2\Fr{\rd^2 \bos{\CS}_\g}{\rd t^\a\rd t^\b}
+\hbar\left(\Fr{\rd \bos{\CS}_\b}{\rd t^\a}\bos{\CS}_\g
   +\bos{\CS}_\b\Fr{\rd \bos{\CS}_\g}{\rd t^\a}
   +\bos{\CS}_\a \Fr{\rd \bos{\CS}_\g}{\rd t^\b}\right) 
   \cr &
- \bos{\CS}_\a\bos{\CS}_\b\bos{\CS}_\g,
}
}
etc. etc.
By taking $\bos{\CK_\CS}$-cohomology class we 
have
\eqn\xxh{
\eqalign{
[\bos{\ma}_n]:\bos{H}^{\otimes n}_{\bos{\CK_\CS}}&\longrightarrow  \bos{H_{\CK_\CS}},
\cr
\biggl[\bos{\ma}_n\left(\bos{\CS}_{\a_1},\ldots,\bos{\CS}_{\a_n}\right)\biggr] &=\bos{\CA}_{\a_1\ldots\a_n}^\g[\bos{\CS}_\g],
}
}
leading to via path integral
\eqn\xxi{
\eqalign{
\bigl<\bos{\ma}_n\bigr>:\bos{H}^{\otimes n}_{\bos{\CK_\CS}}&\longrightarrow  
\Bbbk[[\hbar,(\{t^\a\})]],
\cr
\biggl<\bos{\ma}_n\left(\bos{\CS}_{\a_1},\ldots,\bos{\CS}_{\a_n}\right)\biggr> &=-\hbar\bos{\CA}_{\a_1\ldots\a_n}^\g\bos{\CZ}_\g,
}
}
by path integrals, i.e.,
\eqn\xxj{
\eqalign{
\biggl<\bos{\ma}_n\left(\bos{\CS}_{\a_1},\ldots,\bos{\CS}_{\a_n}\right)\biggr> 
\equiv\Fr{\hbar^n \rd^n \bos{\CZ}}{\rd t^{\a_1}\ldots\rd t^{\a_n}}
\equiv\bos{\CA}_{\a_1\ldots\a_n}^\g\left< \bos{\CS}_\g\right>
\equiv-\hbar\bos{\CA}_{\a_1\ldots\a_n}^\g\bos{\CZ}_\g.
}
}

It follows that
the higher structure functions $\bos{\CA}_{\a_1\ldots\a_n}^\g$
can be determined uniquely in terms of compositions
of $\bos{\CA}^\g_{\a\b}$ and their derivatives;
for an example we have
\eqn\xxk{
\eqalign{
\bos{\CA}_{\a\b\g}^\s =&
-\bos{\CA}^\r_{\b\g} \bos{\CA}^\s_{\a\r}
 +\hbar \Fr{\rd \bos{\CA}_{\b\g}^\s}{\rd t^\a},\cr
 \bos{\CA}_{\a\b\g\r}^\s =&
-\bos{\CA}^\m_{\b\g\r} \bos{\CA}^\s_{\a\m}
 +\hbar \Fr{\rd \bos{\CA}_{\b\g\r}^\s}{\rd t^\a},\cr
 }
}
It is obvious that $\{\bos{A}^\g_{\a_1\ldots\a_n}\}:=
\{\bos{\CA}^\g_{\a_1\ldots\a_n}|_{\vec{t}=\vec{0}}\}
$ are the structure constants of the initial QFT in the previous subsection.

\subsection{What's Next?}

We recall that all those relations described so far
direct consequences of master equation and
the definition of path integrals with certain mild assumptions. 
Note, however, that our crucial
$\bos{\Delta}$ is ill defined and and consequently 
the definition of $\bos{\CZ}$ is not immune.
We may regard all of our endeavor as an effort to
find  certain graded moduli space $\mathfrak{M}$ parametrizing
family of QFTs. Then the "given" QFT is regarded as a
a base point $o$ in the moduli space $\mathfrak{M}$, where we have a quantum flat structure. A  basis $\{\bos{\CS}_\a\}$ of tangent space $T\mathfrak{M}$ 
corresponds to a basis of all observables and the dual basis 
 $\{t^\a\}$ correspond to local coordinates, respectively, both in the neighborhood of
$o \in \mathfrak{M}$. 
Over $\mathfrak{M}$ we consider certain quantum  bundle
$\mathfrak{Q}\rightarrow \mathfrak{M}$ with
a formal power series  graded-connection $\bos{\CD}$
\eqn\xxl{
\bos{\CD} = \Fr{1}{\hbar}\bos{\CD}^{(-1)} + \sum_{n=0}^\infty \hbar^n 
\bos{\CD}^{(n)}
}
which may be written in terms of local coordinates,
\eqn\xxm{
\bos{\CD} = dt^\a\Fr{\rd}{\rd t^\a}\mathbb{I}  +\Fr{1}{\hbar}\bos{\CA}_\a dt^a
}
where $\bos{\CA}_\a =\sum_{n=0}^\infty \hbar^n \bos{\CA}^{(n)}_\a$,
regarding as matrices $(\bos{\CA}_\a)^{\b}{}_\g$. 
Then the relation \bbu\ implies that $\bos{\CD}$ is {\it flat};
\eqn\xxn{
\bos{\CD}^2=0,
}
and the equation \bbq\ implies that $\{\bos{\CZ}_\a\}$, for a choice of homology class
of Lagrangian subspace, is a flat section.

Now it becomes obvious that understanding "global" properties
of the quantum flat bundle $\mathfrak{Q}\rightarrow\mathfrak{M}$
would be crucial. We may regard the base point $o\in \mathfrak{M}$, where we started
our journey, corresponds to a point of regular singularity at the origin
for the system of differential equations \bbq.
The flat connection above may be degenerated at
certain limiting points in $\mathfrak{M}$. Thus
we compactify $\mathfrak{M} \subset \overline\mathfrak{M}$
by adding all the bad points and consider associated completion
$\overline\mathfrak{Q}\rightarrow  \overline\mathfrak{M}$.
In Sect.~$3.2$ we assumed that $\{\bos{\CS}_\a\}$
form a complete basis of $\bos{\CK_\CS}$-cohomology, which
assumption leds to the flatness \xxn. We expect that for
the generic value of $\{t^\a\}$ the above assumption remains
valid, while for certain degenerated limits it would fail.
In such a degenerated point some odd $\bos{\CK_\CS}$-cohomology
classes would appear. Such cohomology classes correpond to
new (extended) gauge symmetry, which should be resolved
as in Sect.~$2.3.2$. After such a resolution we need to
repeat the procedure to find new family of QFTs from the
degenerated point and try to build up "global" moduli space.
Assuming such a procedure can be done for a given (perturbative) QFT we may
interpret other degenerated points as
different perturbative QFTs.\foot{It may not be a far fetched idea 
that the conjectured M-theory moduli space may be studied by, if possible,
searching family of (covariant) superstrings.}

\subsubsection{Semi-Classical Limit of Quantum Flat Structure}

\noindent
Recall that the structure functions $\bos{\CA}^\g_{\a\b}$
is formal power series in $\hbar$, hiding indices;
\eqn\bbw{
\bos{\CA} = \bos{\CA}^{(0)} + \sum_{n=1}^\infty \hbar^n \bos{\CA}^{(n)}
}
where $\bos{\CA}^{(n)}$ for all $n$ are function of $\{t^\a\}$ independent
to $\hbar$. Let us denote for simplicity in notation
\eqn\bca{
\bos{\CA}^{(0)} = \CA,\qquad \bos{\CA}^{(1)}=\CB,\qquad \bos{\CA}^{(2)}=\CC, \quad
\ldots, 
}
etc.
{}From \bbu\ we find the following infinite sequence of  equations
starting the lowest order in $\hbar$
\eqn\bcb{
\eqalign{ 
0=&\CA_{\a\b}^\r \CA_{\r\g}^\s -\CA_{\b\g}^\r \CA_{\r\a}^\s
,\cr
0=&\rd_\a \CA_{\g\b}^\s- \rd_\g \CA_{\a\b}^\s 
\cr
&
+\left(
\CB_{\a\b}^\r \CA_{\r\g}^\s 
+\CA_{\a\b}^\r \CB_{\r\g}^\s 
-\left(\CB_{\b\g}^\r \CA_{\r\a}^\s
+\CA_{\b\g}^\r \CB_{\r\a}^\s\right)
\right)
,\cr
0=&
\rd_\a \CB_{\g\b}^\s- \rd_\g \CB_{\a\b}^\s 
+(-1)^{|\b||\g|}\left(
\CB_{\a\b}^\r \CB_{\r\g}^\s - \CB_{\b\g}^\r \CB_{\r\a}^\s
\right)\cr
&
+\left(
\CA_{\a\b}^\r \CC_{\r\g}^\s 
+\CC_{\a\b}^\r \CA_{\r\g}^\s 
-\left(\CA_{\b\g}^\r \CC_{\r\a}^\s
+\CC_{\b\g}^\r \CA_{\r\a}^\s\right)
\right),
\cr
\vdots =&\vdots
}
}
etc. etc.
{}From \bbqa\ we find the following sequence of  equations
starting the lowest order in $\hbar$
\eqn\bcba{
\eqalign{
\CA_{\a\b}^\s &= 
\CA_{0\a}^\r \CA_{\r\b}^\s,\cr
\CB_{\a\b}^\s &= 
	\CA_{0\a}^\r \CB_{\r\b}^\s
	+\CB_{0\a}^\r \CA_{\r\b}^\s
	-\Fr{\rd \CA_{0\a}^\s}{\rd t^\b} ,\cr
\CC_{\a\b}^\s &= 
	\CB_{0\a}^\r \CB_{\r\b}^\s
	+\CA_{0\a}^\r \CC_{\r\b}^\s
	+\CC_{0\a}^\r \CA_{\r\b}^\s
-2\Fr{\rd \CB_{0\a}^\s}{\rd t^\b} ,\cr
}
}
etc. etc.

Here are some
remarks on the semiclassical 
 $\hbar\rightarrow 0$ limit of
the master equation \bbh;
\eqn\bcc{
\bos{\Delta \CS} =0, \qquad \left(\bos{\CS},\bos{\CS}\right)=0.
}
Then we have
\eqn\xta{
\bos{\CQ}^2_{\bos{\CS}}=0,\qquad
\bos{\Delta}\bos{\CQ_\CS}+\bos{\CQ_\CS}\bos{\Delta}=0,
}
and we can set $\bos{\CS} = \bos{\CS}^{(0)}$
such that both $\bos{\CS}$ and $\bos{\CQ_\CS}$
are independent to $\hbar$.
It also follows that
$\bos{\Delta}\bos{\CS}_\a=\bos{\CQ}_{\bos{\CS}}\bos{\CS}_\a
=0$ and the relation \bbp\ 
implies that
\eqn\xub{
\bos{\CS}_{\a}\bos{\CS}_{\b}
=\bos{\CA}^{(0)\g}_{\a\b}\bos{\CS}_{\g}
+\bos{\CQ}_{\bos{\CS}}\bos{\La}^{(0)}_{\a\b}
}
and
\eqn\xuc{
\eqalign{
\left(\bos{\CA}^{(1)}\right)^{\g}_{\a\b}\bos{\CS}_{\g}
+\bos{\CQ}_{\bos{\CS}}\bos{\La}^{(1)}_{\a\b}
&=
-\Fr{\rd\bos{\CS}_{\b}}{\rd t^\a} 
+\bos{\Delta}\bos{\La}^{(0)}_{\a\b},
}
}
while, for $n\geq 2$
\eqn\xud{
\left(\bos{\CA}^{(n)}\right)^{\g}_{\a\b}\bos{\CS}_{\g}
+\bos{\CQ}_{\bos{\CS}}\bos{\La}^{(n)}_{\a\b}
=\bos{\Delta}\bos{\La}^{(n-1)}_{\a\b}.
}
Note that the relation \xub\  is just classical algebra of $\bos{\CQ}_{\bos{\CS}^{(0)}}$
cohomology classes in the basis $\{\bos{\CS^{(0)}}_\a\}$ under the product.

Assuming the semi-classical master equation \bcc\ holds,
we shall call the coordinates $\{t^\a\}$ or the basis $\{\bos{\CS}_\g\}$
{\it classical}  if the following equations
are satisfied;
\eqn\xue{
\bos{\Delta}\bos{\La}^{(0)}_{\a\b}=\Fr{\rd\bos{\CS}_{\b}}{\rd t^\a}.
}
Combining above with \xub, we have
\eqn\xueg{
\eqalign{
-{\hbar}\Fr{\rd \bos{\CS}_\b}{\rd t^\a}+
\bos{\CS}_{\a}\bos{\CS}_{\b}
&=\bos{\CA}^{(0)\g}_{\a\b}\bos{\CS}_{\g}
-\hbar\bos{\Delta}\bos{\La}^{(0)}_{\a\b}
+\bos{\CQ}_{\bos{\CS}}\bos{\La}^{(0)}_{\a\b}
\cr
&=\bos{\CA}^{(0)\g}_{\a\b}\bos{\CS}_{\g}
+\bos{\CK_\CS}\bos{\La}^{(0)}_{\a\b}
}
}
If follows, from \xuc\ and \xud, that
$\bos{\CA}^{(n)}=0$ for $\forall n\geq 1$ as $\{\bos{\CS}_\g\}$ 
is a basis of $\bos{\CQ_\CS}$-cohomology group.
Assuming the system admit
a classical coordinates, the structure constants $\bos{\CA}_{\a\b}^\g$
of algebra of $\bos{\CK}_{\bos{\CS}}$-cohomology, in such a coordinates,
are the same with the structure constants 
$\CA_{\a\b}^\g:=\bos{\CA}_{\a\b}^{(0)\g}$ of algebra of $\bos{\CQ}_{\bos{\CS}}$-cohomology.

{\bf Question}: Assuming the semi-classical master equation, does 
a classical system of coordinates always exist?

Assume a classical coordinates system exist and
Let $\{t^\a\}$ be such a system of classical coordinates.
Then 
the equations \bbss, \bbu, and \bbqa\  of quantum-flat structure (in the semi-classical
limit)  reduce to the following
relations (remember that the structure constants  does
not depends on $\hbar$ under the present circumstance);
\begin{enumerate}
\item Commutativity
\eqn\bcda{
{\CA}_{\a\b}^\g =  {\CA}_{\b\a}^\g.
}
\item Associativity
\eqn\bcdb{
{\CA}_{\a\b}^\r {\CA}_{\r\g}^\s 
={\CA}_{\b\g}^\r {\CA}_{\r\a}^\s.
}
\item "Potentiality"
\eqn\bce{
\eqalign{
\rd_\a {\CA}_{\g\b}^\s&= \rd_\g {\CA}_{\a\b}^\s. 
}
}

\item Identity
\eqn\bcf{
{\CA}_{\a\b}^\s = {\CA}_{0\a}^\r {\CA}_{\r\b}^\s,
\qquad \Fr{\rd {\CA}_{0\a}^\r}{\rd t^\b}=0.
}

\end{enumerate}
We also note that
$\bos{\CS} =\bos{S} 
+\sum_{n=1}^\infty\Fr{1}{n!} t^{\a_1}\ldots t^{\a_n}
\left(\Fr{\rd^n\bos{\CS}}{\rd t^{\a_1}\ldots\rd t^{\a_n}}\biggl|_{\vec{v}=\vec{0}}\right)
$
while in the classical coordinates
we have \xue\ , i.e.,
$
\Fr{\rd^2\bos{\CS}}{\rd t^\a\rd t^\b}= \bos{\Delta\La}^{(0)}_{\a\b}.
$
Thus we obtain
\eqn\lakm{
\bos{\CS} = \bos{S} + t^\a\bos{O}_\a + \sum_{n=2}^\infty \Fr{1}{n!}t^{\a_1}\ldots t^{\a_n}
\bos{\Delta}\bos{\G}_{\a_1\ldots\a_n},
}
where
\eqn\lakmm{
\eqalign{
\bos{O}_\a &= \bos{\CS}_\a\biggl|_{\vec{t}=\vec{0}},\cr
\bos{\G}_{\a_1\ldots\a_n} &= \Fr{\rd^{n-2}\bos{\La}^{(0)}_{\a_{n-1}\a_n}}{\rd t^{\a_1}
\ldots \rd t^{\a_{n-2}}}\biggl|_{\vec{t}=\vec{0}}
}
}

Now we note that the 
semi-classical flat structure together with existence of classical system
coordinates is closely related with 
the well-known flat or Frobenius structure.
The missing piece is an invariant metric $g_{\a\b}$
satisfying
\eqn\bcfa{
{\CA}_{\a\b}^\r g_{\r\g} ={\CA}_{\b\g}^\r
g_{\r\a},\qquad \Fr{\rd g_{\a\b}}{\rd t^\g}=0.
}
On the other hand we already started from generating functional
$\bos{\CZ}$ for the family of QFTs.

Historically the flat structure was first discovered by K.~Saito
in his study of the period mapping for a universal unfolding of
a function with an isolated critical point (in the context of 
singularity theory) \cite{S}. His main motivation was to have a new
constructions of modular functions by a certain generalization
of the well-known theory of elliptic integrals and modular functions
(see a recent exposition \cite{Sa}).
Here the concept of primitive form, which essentially plays the role
of $\bos{\CZ}$ is crucial. We may
interpret his work as a kind of the rigorous and complete description
of, perhaps, the simplest example of flat family of QFTs, which also
give us a hint on the global issues of flat family of QFTs .
In the physics literature flat structure first appear, independently, in the works
of Witten-Dijkraaf-Verlinde-Verlinde (WDDV) on topological conformal field theory
in $2$-dimensions \cite{W1,DVV}. Later Dubrovin
formalized those structures in the name of Frobenius manifolds \cite{D}.
Another construction of flat or Frobenius structure is due to Barannikov-Kontsevich
on the extended moduli space of complex structures on Calabi-Yau manifold
in the context of B-model \cite{BK}. 
We should remark that the construction of Frobenius structure by Barannikov-Kontsevich
is more general than their original context and closely related with our semi-classical
case. 
All of the above constructions of flat or Frobenius structures can be
viewed as special limits  of the  quantum flat structures of various
family of QFTs

\newsec{Down to Earth}

One may start from
a ghost number $U=0$ function $\bos{S}^{(0)}$ on an
infinite dimensional graded space $\bos{\CT}$ admitting
odd symplectic structure $\bos{\o}$ carrying the ghost number $U=-1$
satisfying the classical master equation
\eqn\daa{
\left(\bos{S}^{(0)},\bos{S}^{(0)}\right) =0,
}
where $\left(\bullet,\bullet\right)$ is the graded Poisson bracket,
carrying the ghost number $U=1$, defined by $\bos{\o}$. 
An usual classical action functional $\bos{s}$, then, corresponds
to the restriction of $\bos{S}^{(0)}$ to a Lagrangian subspace
$\bos{\CL}$ of $\bos{\CT}$. More precisely classical action functional
$\bos{s}$ is typically supported only on certain subspace, which is called
the space of classical fields, of $\bos{\CL}$. The complimentary of
space of classical fields in $\bos{\CL}$ consists of space of
ghosts and their anti-ghosts multiplets, ghosts of ghosts and their
anti-ghost multiplets, etc., etc., depending on the nature of symmetry
and constraints of the classical action functional.

In this section we describe a systematic ways of
constructing the classical BV action functional
for  a large class of $d$-dimensional QFTs related with symplectic $d$-algebras.
The upshot is that for any symplectic $d$-algebra one can associate
QFT on $d$-dimensions. For QFT on $d$-dimensional manfold with
boundary we shall see that there exist a structure of strongly homology $(d-1)$-algebra
on boundary and the structure of symplectic $d$-algebra in the bulk,
which corresponds to the structure on the cohomology of Hochschild complex
of the boundary algebra. This setup may be used to develop an universal
quantization machine of $(d-1)$-algebras.
Also the construction in this section
can be used to "define" differential-topological invariants of low ($d=3,4$)-dimensions
for any symplectic $d=3,4$ algebras.
We remark that this section is an obvious generalization of the
authors previous work \cite{P1}, which was motivated by the seminal paper
\cite{K2} of Kontsevich \cite{K2} as well as the papers \cite{AKSZ,K1,CF}.

\subsec{Closed $s$-Braneoids}

Let $M_{s+1}$ be an $(s+1)$-dimensional oriented and smooth
manifold. Being an $(s+1)$-dimensional QFT, a classical
BV action functional
$\bos{S}^{(0)}$ must  be defined by integration of certain top-form
over $M_{s+1}$. The differential forms on $M_{s+1}$ may
most suitably be described as smooth functions on the total
space $T[1]M_{s+1}$ of twisted by $U=1$ tangent space
to $M_{s+1}$. Without any loss of generality we may pick
a local coordinates system $\{\s^\m\}$, $\m=1,\ldots,s+1$,
on $M_{s+1}$ and assign $U=0$. Then the fiber coordinates
are set to $\{\th^\m\}$ assigned to $U=1$, such that
$U(\s^\m) + U(\th^\m)=1$ and $\th^\m\th^\n = -\th^\n\th^\m$.
Let  $\mathfrak{s}=\bigoplus_{k=0}^{s+1}\mathfrak{s}_k$ be
the $\Z$-graded space of functions on $T[1]M_{s+1}$. The space
$\mathfrak{s}$ is isomorphic to the space of differential forms
on $M_{s+1}$, where the wedge product is replaced with ordinary
product.
The exterior derivative $d$ on $M_{s+1}$ induces an odd nilpotent vector field
$\hat{d}=\vt^\m\Fr{\rd}{\rd \s^\m}$ carrying $U=1$ on $T[1]M_{s+1}$. 
Thus we have the following
complex
\eqn\dab{
\left(\hat{d}; \mathfrak{s}=\bigoplus_{k=0}^{s+1}\mathfrak{s}_k\right),
}
which is isomorphic to de Rham complex on $M_{s+1}$.
Now the integration of certain $(s+1)$-from over $M_{s+1}$
is equivalent to
\eqn\dac{
\oint_{T[1]M_{s+1}}\a:=
\int_{M_{s+1}}d\s^1\ldots d\s^{s+1}\int d\th^1\ldots d\th^{s+1} \a
}
where $\a\in \ms_{s+1}$, which may take certain values of extra structures.
We note that the integration over odd variable (Berezin integral) is defined
as $\int d\th^\m \th^\n = \d^{\m\n}$. Thus the integral  $\int d\th^1\ldots d\th^{s+1}$
shifts $U$ by $-(s+1)$. We should also note that the total integration
measure is coordinates independent as the Jacobians of even and odd
variable cancel each others.

Now the infinite dimensional graded space $\bos{\CT}$ may be
viewed as space of certain functions on $T[1]M_{s+1}$, more
precisely, the space of all sections of certain graded bundle over $T[1]M_{s+1}$.
On such a space $\bos{\CT}$ the odd symplectic form $\bos{\o}$ must
be induced from something. A natural choice
is to identify $\bos{\CT}$ with the space of all ghost number and parity preserving maps 
\eqn\dba{
\Phi: T[1]M_{s+1}\rightarrow \T_{s+1},
}
where $\T_{s+1}$ is a finite dimensional smooth graded
space admitting symplectic form $\o_s$ of ghost number $U=s$
and of the same parity, even or odd, with $s$.


We denote $\mt$ the space of functions on $\T_{s+1}$, graded
by the ghost number $U$,
\eqn\dad{
\mt =\bigoplus_{k \in \Z} \mt_k.
}
Associated with $\o_s$ we have the graded Poisson
bracket $[\star,\star]_{-s}$ of degree $U=-s$
\eqn\dae{
[\star,\star]_s: \mt_{k_1}\otimes \mt_{k_2} \longrightarrow \mt_{k_1+k_2 -s}.
}
We call the algebra functions on $\T_s$
$(\mt, [\star,\star]_s,\cdot)$
endowed with $[\star,\star]_{-s}$ and ordinary (super) commutative and (super) associative product $\cdot$ a {\it symplectic  $(s+1)$-algebra}.
We remark that the notion of $d=(s+1)$-algebra has been first introduced by
Getzler-Jones \cite{GJ} and refined by  Tarmarkin and Kontsevich (see ref.\ \cite{K2}
for some history and operadic viewpoint, which we shall not use directly here).
We put symplectic as the bracket 
$[\star,\star]_{-s}$ is non-degenerated.

Now we consider an element $H_o \in \mt_{s+1}$ satisfying
\eqn\daf{
[H_o, H_o]_{-s} =0.
}
Define $Q_{H_o}$ as the Hamiltonian vector field on $\T_{s+1}$
\eqn\dag{
Q_{H_o} = [H_o,\ldots]_{-s}.
}
Note that $Q_{H_o}^2=0$ and $Q_{H_o}$ is odd carrying $U=1$.
Thus we have a complex
\eqn\dah{
\left(Q_{H_o}, \mt =\bigoplus_k \mt_k\right),
}
to be compared with \dab.

Remark that $[\star,\star]_{-s}:\mt_{s}\otimes \mt_{s}\rightarrow \mt_{s}$,
which means that $[\star,\star]_{-s}$ induces a structure of Lie algebra
on $\mt_{s}$. Thus we may define adjoint adjoint action
by an element $b \in \mathfrak{t}_s$ an any $\g \in \mt$;
\eqn\dbb{
e^{ad_b}\circ (\g) := \g + [b,\g]_{-s} +{1\over 2!}[b,[b,\g]_{-s}]_{-s} +
\ldots,
}
which  is equivalent to a degree
preserving canonical transformation connected to the identity.
We call the two solutions $H, H^\prime \in \mt_{s+1}$ of \daf\ equivalent
if they are related by the above adjoint action.
Thus we can define a moduli space $\CN$ as the set of equivalence
classes of solutions of \daf;
\eqn\dbc{
\CN = \{ H \in \mt_{s+1}| [H, H]_{-s} =0\}/\sim,
}
such that the solution $H_o$ corresponds to a point $o \in \CN$.
Here we consider the case that the ghost number $U=s$ part of $Q_{H_o}$
cohomology of the complex \dah\ is trivial. Otherwise we assume
small resolution of the pair $(\T_{s+1}, H_o)$ to $(\tilde\T_{s+1}, \tilde H_o)$
following the similar procedure described in Sect. $2.3$.
We shall also assume the "anomaly free" condition that 
the ghost number $U=s+2$ part of $Q_{H_o}$
cohomology of the complex \dah\ is trivial. Then the moduli space
$\CN$ is unobstructed (around $o$).

We describe a map \dba\ locally by a "local" coordinates
on $\T_{s+1}$, which are regarded as functions on $T[1]M_{s+1}$.
Let $\{x^I\}$ be a "local" coordinates system on $\T_{s+1}$.
We denotes the ghost number of $x^I$ by $U(x^I)\in \Z$. 
We parametrize
a map by $\{\hat{x}^I\}$, where
\eqn\dai{
\eqalign{
\hat{x}^I &:= x(\s,\th)^I=x(\s)^I
+ \Fr{1}{n!}\sum_{i=1}^{s+1}x(\s)^I_{\m_1\ldots 
\m_{s+1}}\th^{\m_1}\ldots
\th^{\m_{s+1}}.\cr
}
}
By the ghost number preserving maps we meant
\eqn\daj{
U(x^I) = U(\hat{x}^I),\qquad |x^I| =|\hat x^I|.
}
We define associated
$n$-form components on $M_{s+1}$ by
\eqn\dak{
x^I_{[n]} := \Fr{1}{n!} x(\s)^I_{\m_1\ldots
\m_n}d\s^{\m_1}\wedge\ldots\wedge
d\s^{\m_n}.
}
Note that the ghost number $U$ of $x^I_{(n)}$ is $U(x^I) -n$.
{}From the symplectic form $\o_s(dx, dx)$ of $U=s$ we have
the following induced odd symplectic form $\bos{\o}$ carrying $U=-1$
on $\bos{\CT}$
\eqn\dbd{
\bos{\o} = \oint_{T[1]M_{s+1}}\o_s(\d \hat x, \d \hat x),
}
since $\oint_{T[1]M_{s+1}}$ shifts $U$ by $-(s+1)$
and has the parity $(s+1)\hbox{ mod } 2$.

Now we can describe  classical BV action functional
associated to $H_o$
as follows
\eqn\dal{
\eqalign{
\bos{S}^{(0)}_{H_o}
&= \oint_{T[1]M_{s+1}} \left(
\o(\hat{x},{\hat d \hat x})_s + \Phi^*(H_o)\right)
\cr
&\equiv 
\oint_{T[1]M_{s+1}} \left(
\o(\hat{x},\hat{d}\hat{x})_s + \hat {H}_o\right)
}
}
Note that $\bos{S}^{(0)}_{H_o}$ carries the ghost number $U=0$
and is an even function(al).
One can check that $\bos{S}^{(0)}_{H_o}$ satisfies the
classical master equation $\left(\bos{S}^{(0)}_{H_o},\bos{S}^{(0)}_{H_o}\right)=0$
provided that  the boundary of $M_{s+1}$ is empty, as the result of
$[H_o,H_o]_{-s}=0$,
after integration by parts and using the Stokes theorem.

Define odd Hamtionian vector of $\bos{S}^{(0)}_{H_o}$
on $\bos{\CT}$ as follows;
\eqn\dam{
\bos{Q}_{\bos{S}_{H_o}^{(0)}} :=\left(\bos{S}_{H_o}^{(0)},\ldots\right),\qquad 
\bos{Q}^2_{\bos{S}_{H_o}^{(0)}} =0,
}
and by definition $\bos{Q}_{\bos{S}_{H_o}^{(0)}}\bos{S}_{H_o}^{(0)}=0$.
We note that
\eqn\dama{
\bos{Q}_{\bos{S}_{H_o}^{(0)}}\hat x^I = \hat d\hat x^I + \widehat{Q_{H_o} x^I}.
}
Since $\bos{Q}_{\bos{S}^{(0)}_{H_o}}^2=0$, we have the following
complex
\eqn\daw{
\left(\bos{Q}_{\bos{S}^{(0)}_{H_o}}, \mT =
\bigoplus_{k\in \Z}\mT_k\right).
}
An classical observable of the theory is an element of cohomology
of the above complex. A class of classica observables  can be constructed as follows.

Consider a function $\g \in \mt$ on $\T_s$ with certain ghost number $U(g)$.
The pullback $\hat{\g}:=\bos{\Phi}^*(\g)$ of $\g$
can be viewed as a functional $\g(\s^\m,\th^\n)$ on $T[1]M_{s+1}$.
The relation \dama\ implies that 
$\bos{Q}_{\bos{S}^{(0)}_{H_o}} \hat{\g} = \hat{d} \hat{\g} + \hat{Q_{H_o}\g}$.
Let  $\g$ be an elment of the $Q_{H_o}$-cohomology of the complex \dah.
Then  we have
\eqn\day{
\bos{Q}_{\bos{S}^{(0)}_{H_o}} \hat{\g} = \hat{d} \hat{\g},
}
for $\g \in H(\mt,Q_{H_o})$.
We call the above relation {\it descent equations}.
We can expand $\hat{\g}$ as
\eqn\daz{
\eqalign{
\hat{\g} =& \g(\s)
+ \sum_{n=1}^{s+1}\g(\s)_{\m_1\ldots\m_n}\th^{\m_1}\ldots\th^{\m_n},
}
}
where $\g(\s)_{\m_1\ldots\m_n}$ is a functional on $M_{s+1}$
transforming as totally antisymmetric $n$-tensor.
Then we have associated $n$-form $\g_{[n]}$ on $M_{s+1}$
defined by
\eqn\dca{
\g_{[n]} := \Fr{1}{n!} \g(\s)_{\m_1\ldots
\m_n}d\s^{\m_1}\wedge\ldots\wedge
d\s^{\m_n}.
}
Note that the ghost number $U$ of $\g_{[n]}$ is $U(\g) -n$.
Then the equation \day\ becomes
\eqn\dcb{
\eqalign{
\bos{Q}_{\bos{S}^{(0)}_{H_o}}\g_{[0]} =0,\cr
d\g_{[0]}- \bos{Q}_{\bos{S}^{(0)}_{H_o}}\g_{[1]}=0,\cr
d\g_{[1]}- \bos{Q}_{\bos{S}^{(0)}_{H_o}}\g_{[2]}=0,\cr
\vdots,\cr
d\g_{[s]}- \bos{Q}_{\bos{S}^{(0)}_{H_o}}\g_{[s+1]}=0,\cr
d\g_{[s+1]}=0,
}
}
where $d$ denotes the exterior derivative on $M_{s+1}$.
Now consider a homology $n$-cycle $C_n \in H_*(M_{s+1})$ on $M_{s+1}$
and define $\int_{C_n} \g_{[n]}$. Then \dae\ implies
that $\int_{C_n}\g^{[n]}$ is an observable
and the BRST cohomology of $\bos{Q}_{\bos{S}^{(0)}_{H_o}}$ depends only on the
homology class of $C_n \in H_*(M_{s+1})$.

Now we {\it assume} that one can define $\bos{\Delta}$
after suitable regularization such that $\bos{\Delta}^2=0$.
We shall consider the situation that there exist $\bos{S}_{H_o}$
satisfying
\eqn\dcc{
-\hbar\bos{\Delta}\bos{S}_{H_o} +\Fr{1}{2}\left(\bos{S}_{H_o},
\bos{S}_{H_o}\right)=0,
}
where 
\eqn\dcd{
\bos{S}_{H_o} =\bos{S}^{(0)}_{H_o} + \hbar \bos{S}^{(1)}_{H_o}
+\ldots.
}

Now we back to the target space $\T_{s+1}$.
Let $\{\g_\a\}$ be a basis of the cohomology of the complex
\dah\ and let $\{t^\a\}$ be the dual basis such
that
\eqn\dce{
U(\g_\a) + U(t^\a)=s+1.
}
Now we assume that there is solution of 
\eqn\dcf{
[\CH, \CH]_{-s}=0,
}
such that
\eqn\dcg{
\CH = H + t^\a\g_\a + \sum_{n=2}^\infty \Fr{1}{n!}t^{\a_1}\ldots t^{\a_n}\g_{\a_1\ldots\a_n},
\qquad U(\CH) =s+1.
}
Let's define $\CQ_\CH =[\CH,\ldots]_{-s}$ which is
odd nilpotent vector of degree $U=1$.

In general we consider certain graded Artin ring with
maximal ideal
\eqn\dch{
\mathfrak{a} = \bigoplus_{-\infty < k \leq s+1}\mathfrak{a}_k.
}
The bracket $[\star,\star]_{-s}$ on $\mt$ can be naturally
extended to $\mt\otimes \ma$. Then we may define
extended moduli space $\CM$;
\eqn\dci{
\CM = \{\CH \in (\mt\otimes \ma)_{s+1}|[\CH,\CH]_{-s}=0\}/\sim
}
where the equivalence is defined by adjoint action
of element $\b \in (\mt\otimes \ma)_s$.
Then we may regard the fixed $H \in \CN\subset \CM$
as a basepoint $o$ in $\CM$ and interpret $Q_H$-cohomology
as the tangent space $T_o\CM$.

Now we have corresponding families of classical BV action functional
\eqn\dcj{
\eqalign{
\bos{\CS}^{(0)}_\CH 
&= \int_{T[1]M_{s+1}} \left(
\o(\hat{x},\hat{d}\hat{x})_s + \hat {\CH}\right)
\cr
&= \bos{S}^{(0)}_H + 
 \int_{T[1]M_{s+1}}\left(
t^\a\hat \g_\a + \sum_{n=2}^\infty \Fr{1}{n!}t^{\a_1}\ldots t^{\a_n}\hat\g_{\a_1\ldots\a_n},
\right)
}
}
{}From \dcf, and as the boundary of $M_{s+1}$ is empty, we have
\eqn\dck{
\left(\bos{\CS}^{(0)}_\CH,\bos{\CS}^{(0)}_\CH\right)=0,
}
Thus
\eqn\dcl{
\bos{\CQ}_{\bos{\CS}_\CH^{(0)}}=\left(\bos{\CS}_\CH^{(0)},\ldots\right),
\qquad
\bos{\CQ}^2_{\CS_\CH^{(0)}}=0,
\qquad 
\bos{\CQ}_{\bos{\CS}_\CH^{(0)}}\hat x^I = \hat d\hat x^I + \widehat{\CQ_\CH x^I}.
}

Thus we have
\begin{quote}
{\it
Associated with any symplectic $(s+1)$-algebra with non-empty
moduli space $\CM$
we have family of pre QFTs which
action functional satisfies the classical BV master equations.
}

\end{quote}

Now we {\it assume} that one can define $\bos{\Delta}$
after suitable regularization such that $\bos{\Delta}^2=0$.
We shall consider the situation that for any  $\bos{\CS}^{(0)}_\CH$,
$\CH \in \mathfrak{M}$ there exist $\bos{\CS}_{\CH}$
satisfying
\eqn\dcm{
-\hbar\bos{\Delta}\bos{\CS}_\CH +\Fr{1}{2}\left(\bos{\CS}_\CH,
\bos{\CS}_\CH\right)=0,
}
where 
\eqn\dcn{
\bos{\CS}_\CH =\bos{\CS}^{(0)}_\CH + \hbar \bos{\CS}^{(1)}_\CH
+\ldots.
}
Then we have
\begin{quote}
{\it
Family of QFTs parametrized by 
the moduli space $\mathfrak{M}$.}

\end{quote}
Now our earlier discussion endows  quantum flat structure on $\mathfrak{M}$
via the family of $(s+1)$-dimensional QFTs, which define
the function $\bos{\CZ}$ on $\mathfrak{M}$;
\eqn\dco{
\bos{\CZ} = \int_{\bos{\CL}}d\m\, e^{-\bos{\CS}_\CH/\hbar}
}
Recall that $\bos{\CZ}$ depends on homology classes of the
Lagrangian subspace $\CL$ in $\bos{\CT}$. So we have any many
inequivalent flat structures on $\mathfrak{N}$ as homology classes
of Lagrangian subspaces. 

We also note that, by construction,
\begin{quote}
{\it
The quantum flat structures on $\mathfrak{N}$ 
depend on the smooth structures on $M_{s+1}$.
}
\end{quote}
It can be argued that
the following is true.
\begin{quote}
{\it
There exist a suitable choice of homology class $[\bos{\CL}]$ of
Lagrangian subspaces and its
representative  such that $\bos{\CZ}$ define
family of differential-topological invariants on $M_{s+1}$.
}
\end{quote}
The most interesting case  for the above perspective would be
$s=3$, that is, quantum field theoretic definition of smooth invariants of 
$4$-manifold.\foot{Allow me to give a simple example for this
case. Let $V$ is a finite dimesional vector space over $R$
and let $V[1]$ be the suspension of $V$ by $U=1$. Then
consider $\T_4 = T^*[3]V[1] \simeq V[1]\oplus V^*[2]$,
which has a structure of symplectic $4$-algebra.
The resulting $3$-braneoid leads to the celabrated 
Donaldson-Witten theory \cite{D,W}
after suitable gauge fixing for $dim V < 4$. In general
we have certain deformation of Donaldson-Witten theory,
where semi-simple Lie algebra is replaced with semi-simple
weakly homotopy Lie algebra, where the Jacobi identity is violated.
Nonetheless the path integral gives smooth invariants of
$4$-manifolds. The similar deformation is also possible
for physical Yang-Mills theory.
}
Note that the Hodge star operator $*$ on smooth oriented $4$-manifold
satisfies $*^2=1$ and maps $2$-form to $2$-form. This implies
that we can always choose a contiuos family of Lagrangian subspaces $\bos{\CL}$
depending on continuous family of metric, which property implies
that $\bos{\CZ}$ define
family of smooth invariants of $4$-manifold. The philosophy
here is to use all symplectic $4$-algebras to prove
smooth structures of $4$-manifold via associated QFTs.

We remark that the whole construction described above can
be generalized by replacing $T[1]M_{s+1}$ with any smooth graded
manifold adimiting a non-degenerated volume form of the parity of $(s+1)$
with $U=-(s+1)$ and an odd nilpotent vector field with $U=1$.

\subsubsection{Hamitonian picture and dimensional reduction}

Now we turn to Hamiltonian picture and dimensional reduction.

Consider the classical BV action functional $\bos{S}^{(0)}_{H_o}$ in
\dal. It is not difficult to see that $\bos{S}^{(0)}_{H_o}$ is invariant
under odd symmetry generated by $\bos{V}_\m$ carrying $U=-1$;
\eqn\dcp{
\bos{V}_\m \hat x^I := \Fr{\rd}{\rd \th^\m}\hat x^I,
}
as $\bos{S}^{(0)}_{H_o}$ is defined by an integral over $T[1]M_{s+1}$.
We have the following commutation relation
\eqn\dcq{
\{\bos{V}_\m,\bos{V}_\n\}=0,\qquad \{\bos{V}_\m,\bos{Q}_{\bos{S}_{H_o}^{(0)}}\}
= \Fr{\rd}{\rd \s^\m},
}
which is a form of world-volume supersymmetry.
Now we assume that $M_{s+1}= M_s \times \R$, where
$M_s$ is an oriented smooth $s$-dimensional manifold.
We can decompose $(\s^\m)$ as $(\s^i, \s^0)$, $i=1,2,\ldots, s$,
where $s^0$ is the time $(\R)$ coordinates. Then the component
$\bos{V}_0$ is defined globally as $\bos{Q}_{\bos{S}_{H_o}^{(0)}}$.
Note that both $\bos{V}_0$ and $\bos{Q}_{\bos{S}_{H_o}^{(0)}}$
are odd nilpotent vector fields on $\bos{\CT}$. Let
$\bos{Q}^*$ and $\bos{Q}$ denote corresponding charges.
Then the commutation relation 
$\{\bos{V}_0,\bos{Q}_{\bos{S}_{H_o}^{(0)}}\}
= \Fr{\rd}{\rd \s^0}$ implies that
\eqn\cdr{
\bos{Q}^*\bos{Q} +\bos{Q}\bos{Q}^* = Ham
}
where $Ham$ means "Hamiltonian" of the theory (the usual
Hamiltonian can be obtained from $Ham$ after "gauge fixing") .
We may interested in the ground state $Ham|0>=0$.\foot{Applying
this construction to the case in the foonote$^{12}$ leads to the
Floer homology of $3$-manifolds \cite{Fl,W} for
$dim V <4$ and its deformation in general.
This implies Floer-like homology of $3$-manifolds
has a full featured generalization associated with symplecitc
$4$-algebras.
}

The dimensional reduction means dropping the dependence 
of the theory on the "time-direction" $\R$. Note that
the superfields $\hat x^I =x(\s^m, \th^\m)$ are decomposed
as
\eqn\dcs{
\hat x^I =x^I(\s^i, \th^i,\th^0) = y^I(\s^i,\th^i) + \th^0 z^I(\s^i,\th^i)
 := \hat y^I + \th^0 \hat z^I
}
such that $U(\hat y^I) =U(\hat x^I)$ and $U(\hat z^I) = U(\hat x^I)-1$.
It follows that the dimensional reduction of the theory
of the maps $T[1]M_{s+1}\rightarrow \T_{s+1}$ becomes
a theory of maps $T[1]M_{s}\rightarrow T^*[s-1]\T_{s+1}$.\foot{Applying
this construction to the case in the foonote$^{12}$ leads to the
Casson invariant of $3$-manifold \CW.}

\subsection{Open $s$-Braneoids}

So far we assumed that the $(s+1)$-dimensional manifold
$M_{s+1}$ has no boundary. Now we consider the
cases that the boundaries of $M_{s+1}$ are non-empty.
For simplicity we shall begin with the case that
there is only one boundary component.

\subsubsection{Boundary Condition}

Assume that we have the same data as the empty boundary
case and consider $\bos{S}^{(0)}$ given by \dal.
Recall that the BV bracket $\left(\bos{S}^{(0)},\bos{S}^{(0)}\right)$
involves a total derivative term from the bracket between the
first term $\oint_{T[1]M_{s+1}}\o_s(\hat x, \hat d \hat x)$ 
in \dal. It is not difficult to check the total derivative
term vanish if we impose the following boundary condition
\eqn\dda{
\eqalign{
\Phi&: T[1]M_{s+1}\rightarrow \T_{s+1},\cr
\Phi&(T[1](\rd M_{s+1}) \subset \L,
}
}
where $\L$ is a any Lagrangian subspace of $\T_{s+1}$
with respect to the symplectic form $\o_s$.
We remark that for $s=even$ $\T_{s+1}$ may not
admits any Lagrangian subspace. From now on
we always consider $(\T_{s+1},\o_s)$ admitting Lagrangian
subspace. The BV bracket $\left(\bos{S}^{(0)},\bos{S}^{(0)}\right)$
involves another total derivative term from the bracket between the
first term $\oint_{T[1]M_{s+1}}\o_s(\hat x, \hat d \hat x)$
and the second term $\oint_{T[1]M_{s+1}}\hat H_o$ in \dal. 
With the above boundary
condition such total derivative term vanishes
if
\eqn\ddb{
H|_\L =0.
}
Finally the bracket between the second term in \dal\
vanishes iff
\eqn\ddc{
[H, H]_{-s} =0.
}

\subsubsection{Algebraic Digression}

We may identify $\T_{s+1}$ in the neighborhood of $\L$ with
the total space $T^*[s]\L$ of cotangent bundle over $\L$ with the fiber
twisted by $U=s$;
We  denote a system of  Darboux coordinates of $T^*[s]\L$
by $(\mq^{a}, \mp_{a})$, $(base|fiber)$ such that
\eqn\ddd{
\o_{s} = d\mp_a d\mq^a,\qquad U(\mq^{\a}) +U(\mp_{\a}) = s,
}
and $\L$ is defined by $\mp_a =0$ for all $a$.
Then one may Taylor expand $H\in\mt_{s+1}$ around
$\L$ 
\eqn\dde{
\eqalign{
H &= \sum_{n=0}^\infty M_n,\cr
M_n & 
= \Fr{1}{n!} {m}(\mq)^{a_1\ldots a_n} \mp_{a_1}\ldots  \mp_{a_n}.
}
}
Now the condition $[H,H]_{-s}=0$ becomes
\eqn\ddf{
\sum_{p+q =n}\left[M_p , M_q\right]_{-s}=0,\qquad \forall n \geq 0.
}
For each $M_n$ one may assign $n$-poly differential
operator ${m}_n$ acting on the $n$-th tensor product $\ml^{\otimes m}$ of the space $\ml$ of functions
on $\L$ such as; 
\eqn\ddg{
\eqalign{
{m}_0:  &\ml\rightarrow \Bbbk,\cr
{m}_n: &\ml^{\otimes n}\rightarrow \ml,\qquad\hbox{for } n\geq 1
} 
}
by canonically "quantization", i.e., replacing
the BV bracket $(\mp_a, \mq^b)=\d_{a}{}^b$
to commutators of operators,  
naively, $\hat\mq^b = \mq^b$ and $\hat\mp_a = \Fr{\rd}{\rd \mq^a}$.
It is not difficult to check $m_n$ carry ghost number
\eqn\ddh{
U(m_n)=-ns + s+1,
} 
i.e., $U(m_0)=s+1$, $U(m_1)=1$, $U(m_2)= -s+1$,
etc. 
Then we may take the condition \ddf\ as a definition
of a structure $(m_0,m_1,m_2,\ldots)$ of {\it weakly
homotopy Lie $s$-algebroid} on $\L$.
We note that $H|_\L = m_0$. Thus the condition $H|_\L=0$
means that $m_0=0$ and together with the condition
$[H,H]_{-s}=0$ we have a a structure  $(m_1,m_2,\ldots)$ of {\it strongly
homotopy Lie $s$-algebroid} on $\L$.\foot{A structure of strongly
homotopy Lie $s$-algebroid is certain homotopy generalization of
structure of Lie algebroid on $\L$. Note that the equation \ddf\ with
the condition $M_0=0$ reads
\eqn\ddi{
\eqalign{
[M_1,M_1]_{-s}=0,\cr
[M_1,M_2]_{-s}=0,\cr
\Fr{1}{2}[M_2,M_2]_{-s} + [M_1,M_3]_{-s}=0,\cr
[M_2,M_3]_{-s} + [M_1,M_4]_{-s}=0,\cr
\vdots \qquad
}
}
etc.
Note that $m_1 = m(q)^a\Fr{\rd}{\rd q^a}$
is an odd vector with $U=1$ on $\L$, which satisfies
$m_1^2=0$, due to the first equation above. We
may take the existence of such odd vector field 
as a structure of 
of Lie algebroid on the graded space $\L$. 
The second equation above may be viewed as
the condition that $M_2$ defines a cocycle.
Assume that $M_n=0$ for $\forall n\geq 3$, we may call
above as a structure of $s$-Lie bi-algebroid. For
$M_n=0$ for $\forall n\geq 4$, we may call above as a structure
of quasi $s$-Lie bi-algebroid etc. Some important examples
of such structure are in refs.\ \cite{Dr2,RoyW,P1,HP1,OP1}.
}

We call two structures of 
strongly homotopy Lie $s$-algebroids on $\L$ are equivalent
if they are related by change of the Lagrangian compliment  of $\L$ in $\T_{s+1}\simeq
T^*[s]\L$,
called skrooching.  It is obvious a skrooching always leads to
another structure of strongly homotopy Lie $s$-algebroids on $\L$
as a skrooching always preserve all the conditions in \ddb.

Now we consider
the following infinitesimal
canonical transformations
\eqn\ddj{
\eqalign{
\mq^a &\rightarrow \mq^a,\cr
\mp_a &\rightarrow \mp_a + \Fr{\rd \G}{\rd \mq^a},
}
}
generated by $\G(\mq) \in \ml_s$.
Let $H_\G$ be the result of the above transformation,
which automatically satisfies $[H_\G, H_\G]_{-s}=0$,
while $H_\G|_\L \neq 0$ in general; we have
\eqn\ddk{
\eqalign{
H_\G &= \sum_0^\infty M_{\G n},\cr
M_{\G0} &= \sum_{n=1}^\infty \Fr{1}{n!} m_n(\G,\ldots,\G),\cr 
M_{\G 1} &=\sum_{n=1}^\infty\Fr{1}{(n-1)!} 
m^{a_1\ldots a_n}\left(\Fr{\rd \G}{\rd \mq^{a_1}}\right)\cdots\left(\Fr{\rd \G}{\rd \mq^{a_{n-1}}}\right)\mp_{a_n},\cr
&\vdots
}
}

Consider the generating functional $\G\in \ml_s$ satisfying $H_\G|_\L\equiv M_{\G0}\equiv m_{{}_{\G0}}=0$;
\eqn\ddl{
m_1(\G) +\Fr{1}{2}m_2(\G,\G) + \Fr{1}{3!}m_3(\G,\G,\G) +\ldots =0.
} 
Then $H_\G$ induces another structure 
$(m_{{}_{\G1}},m_{{}_{\G2}},m_{{}_{\G3}},\ldots)$
of
strongly homotopy Lie $s$-algebroid on $\L$.
We remark that the two structures $(m_1,m_2,m_3,\ldots)$
and $(m_{{}_{\G1}},m_{{}_{\G2}},m_{{}_{\G3}},\ldots)$ of sh Lie $s$-algebroids
on $\L$ are {\it not} equivalent. We note, as an example, that
\eqn\ddm{
m_{{}_{\G1}} = m_1 + \sum_{n=2}^\infty \Fr{1}{(n-1)!}m_n(\G,\ldots,\G,\phantom{\G})
}
and, by construction, $m_{{}_{\G1}}^2=0$, which is equivalent to the equation \ddk.
We call a solution $\G\in \ml_s$ of the equation \ddk\   strongly homotopy
Dirac $s$-structure. As $\rd_\G:\ml_k\rightarrow \ml_{k+1}$ and $\rd_\G^2=0$ we have
complex $(\rd_\G,\ml)$ and associated non-linear (co)homology by
$"\Ker \rd_\G/\Im \rd_\G"$.

Now we specialize to the case that $s$ is odd. Then the structure
of symplectic $(s+1)$-algebra on $T^*[s]\L$ enhances to
the structure of BV $(s+1)$-algebra. Namely
there exist $\Delta: \mt_k \rightarrow \mt_{k-s}$ satsifying
$\Delta^2=0$;
\eqn\bvs{
\Delta = (-1)^{|q^a|+1}\Fr{\rd_r^2}{\rd q^a \rd p_a},
}
and generate the bracket $[.,.]_{-s}$. For $s=-1$ this
is (finite dimensional version) of the original BV structure in Sect.~$2$.
We remark that the above $\Delta$ should {\it not} be confused
with the BV operator $\bos{\Delta}$ in the space of all fields $\bos{T}$.

This is the end of the digression and let's justify why the above considerations
are relevant.

\subsubsection{Boundary Deformations}

Consider the classical BV action functional
\eqn\ddn{
\bos{\CS}^{(0)}_H 
=\oint_{T[1]M_{s+1}}\left( \hat p^a\hat d\hat q^a + H(\hat q,\hat p)\right),
}
where we assume boundary condition \dda\ and
$H\in \mt_{s+1}$ satisfies
\eqn\ddo{
[H,H]_{-s}=0,\qquad H|_\L=0.
}
Then we may rewrite \ddn\ as follows
\eqn\ddp{
\bos{\CS}^{(0)}_H 
=\oint_{T[1]M_{s+1}}\left( \hat p^a\hat d\hat q^a 
+ \sum_{n=1}^\infty\Fr{1}{n!}m(\hat q)^{a_1\ldots a_n}\hat p_{a_1}\cdots\hat p_{a_n}  \right).
}

Let's now consider canonical transformation generated by 
\eqn\ddq{
\bos{\Psi} = \oint_{M_{s+1}}  \Psi(\hat q,\hat p),
}
where $\Psi(q,p) \in \mt_{s}$ such that $\bos{\Psi} \in \mT_{-1}$.
Let $\G(q) =\Psi(q,p)|_\L \in \ml_{-s}$. The action functional
$\bos{\CS}_{H_o}^{\bos{\Psi}}$ after the resulting canonical transformation
is given by
\eqn\ddr{
\eqalign{
\bos{\CS}_{H}^{(0)\bos{\Psi}} 
=&\oint_{T[1]M_{s+1}} \biggl(\hat{p}_a\hat{d}\hat{q}^a 
+ H_\Psi\left(\hat q, \hat p\right)
 \biggr)
+ \oint_{T[1]M_{s+1}} \hat d \Psi(\hat q,\hat p)
\cr
=&\bos{\CS}_{H_\Psi} + \oint_{T[1](\rd M_{s+1})}\G(\hat q),
}
}
where we used the boundary condition after using the
Stokes theorem. The above action functional
also satisfy the master equation if and only if $H_\Psi|_\L=0$.
On the other hand the value $H_\Psi|_\L$ equals to
$H_\G|_\L$. Thus we have $H_\G|_\L=0$ for the master
equation. We also note that the canonical transformation
generated a boundary interaction term depending only
$\G =\Psi|_\L$.
Thus it is obvious that $\Psi \in \mt_s$ satisfying $\Psi|_\L =0$
leads to the same physical theory. We called canonical transformation
generated by such a $\Psi$ skrooching. On the other hand
$\G \in \ml_s$ leads to a non-zero boundary interaction
terms. According to our definition 
an element $\G \in \ml_s$ satisfying $H_\G|_\L=0$
is a strongly homotopy Dirac $s$-structure on $\L$,
which is defined as a solution of Maurer-Cartan equation \ddl\
of the structure of  strongly homotopy Lie $s$-algebroid, defined
by $H$, on $\L$. Thus the set of equivalence classes of
strongly homology Dirac $s$-structure on $\L$ is isomorphic
to the moduli space of boundary deformations for the fixed
bulk background $H \in \mt_{s+1}$.

\subsubsection{Extended Bulk/Boundary Deformations}

Now we consider bulk deformations compatible with boundary condition.
Consider a deformation $\CH$ \dcg\ of $H$ satisfying
the following equation
\eqn\dea{
\left\{
\eqalign{
[\CH,\CH]=0,\cr
 \CH|_\L=0.
}\right. ,
}
where $\CH\in (\mt\otimes\ma)_{s+1}$.
Thus $\CH$ satisfying above induce a structure
of {\it extended} strongly homotopy Lie $s$-algebroid 
$(\ml, (\m_1,\m_2,\m_3,\ldots))$ on $\L$.
Note that $\m_1^2 =0$. As $\m_1:(\ml\otimes\ma)_k\rightarrow (\ml\otimes\ma)_{k+1}$ and $\m_1^2=0$ we have
complex $(\m_1,\ml)$ and associated (co)homology by
$"\Ker \m_1/\Im \m_1"$. 
We define the extended bulk moduli space
$\mathfrak{N}(\L)_{s+1}$ by the set of solutions of \dea\ modulo
equivalence, defined by the adjoint action
of an element in $(\mt\otimes\ma)_1$ vanishing
on the Lagrangian subspace $\L$ in $\T_{s+1}\simeq T^*[s]\L$ -
this may be called extended skrooching.
We call two structures of 
extended strongly homotopy Lie $s$-algebroids on $\L$ are equivalent
if they are related by extended skrooching. 
Thus $\mathfrak{N}(\L)_{s+1}$ parametrize the set of equivalence
classes of structures of extended sh Lie $s$-algebroid
on $\L$.

It is now natural to consider extended boundary interactions
via extend sh Dirac $1$-structure on $\L$ defined
by elements $\Upsilon\in (\ml\otimes \ma)_s$
satisfying 
\eqn\deb{
\CH_\Upsilon|_\L =0.
}
Equivalently
\eqn\dec{
\m_1(\Upsilon) +\Fr{1}{2}\m_2(\Upsilon,\Upsilon) + \Fr{1}{3!}\m_3(\Upsilon,\Upsilon,\Upsilon) +\ldots =0.
} 
Then $\mH_\Upsilon$ induces another quantizable structure 
$(\m_{{}_{\Upsilon1}},\m_{{}_{\Upsilon{2}}},\m_{{}_{\Upsilon{3}}},\ldots)$
of
strongly homotopy Lie $1$-algebroid on $\L$,
where
\eqn\ded{
\md_\Upsilon = \md + \sum_{n=2}^\infty \Fr{1}{(n-1)!}\m_n(\Upsilon,\ldots,\Upsilon,\phantom{\G})
}
and, by construction, $\md_\Upsilon^2=0$, which is equivalent to the equation \dec.
The two structures $(\md,\m_2,\m_3,\ldots)$
and $(\md_\Upsilon,\m^\Upsilon_{2},\m^\Upsilon_{2},\ldots)$ of extended 
sh Lie $1$-algebroids
on $\L$ are not equivalent. 
As $\md_\Upsilon:(\ml\otimes\ma)_k\rightarrow (\ml\otimes\ma)_{k+1}$ and 
$\md_\Upsilon^2=0$ we have
complex $(\md_\Upsilon,(\ml\otimes\ma))$ and associated non-linear (co)homology by
$"\Ker \rd_\G/\Im \rd_\G"$.

It is obvious that the boundary interaction depend on
the bulk background or the bulk moduli space $\CM_\L$.
Thus the total moduli space $\mathfrak{F}_\L$ of both the bulk and boundary deformations
has a structure of fibered space $\mathfrak{F}_\L\rightarrow \mathfrak{M}_\L$
such that
\eqn\dee{
\matrix{\mathfrak{B}_t&\subset&\mathfrak{F}_\L \cr
\uparrow & &\downarrow \cr
\{t\}&\in &\mathfrak{M}_\L 
}
}

\subsubsection{Deformation Quantization of $d$-Algebra}

Our program is closely related with the  deformation theory of $d$- 
algebra \cite{K2}.
Recently there has been many spectacular developments in deformation
theory of associative algebras \cite{G} (or $1$-algebras from now on),
following the first solution of deformation quantization  
by Kontsevich \cite{K1}.
An amusing  result is that deformation theory of $1$-algebras is
closely related with the geometry of configuration space of points
on $2$-dimensions. Another beautiful result is that the deformation
complex, the Hochschild complex $\Hoch(A_1)$,  of $1$-algebra ($d$-algebra in
general) $A_1$ has
a structure of $2$-algebra ($(d+1)$-algebra) 
\cite{K2,T1}.

Let $\L$ be any $\Z$-graded smooth algebraic variety.
Let $A_s(\L)$ be the algebra of functions on $\L$ regarded as an $s$-algebra.
Let $\Hoch(A_s(\L))$ be the Hochschild complex of $A_s(\L)$
and let $H^\bullet (\Hoch(A_s(\L)))$ be the cohomology. An important
lemma of Kontsevich is that the space $\oplus H^\bullet(\Hoch(A_s(\L))$ is isomorphic
to the space $\mt$ of functions on the total space $\T_{s+1} =T^*[s]\L$
of twisted by $U=[s]$ cotangent bundle to $\L$.  

It is natural to identify  a degree $(s+1)$-function
$H^{s+1}$ on $T^*[s]\L$ with $H|_{\L_s}=0$ as
an element of $H^1(\Hoch(A_s(\L)))$.
The first cohomology of any Hochschild complex of an algebra
is naturally corresponds to the infinitesimal determining the first order
deformation of the algebra. In the present case our $H^{s+1}$
satisfying the "master" equations \aag\ can be interpreted
as the infinitesimal for deformations of  the $s$-algebra $A_s(\L)$ as
an $s$-algebra. This is the first mathematical clue for what kind of 
quantum
algebras  we are dealing with. Our approach also gives a natural
"explanation" why deformation theory of $s$-algebra is related
with differential-geometry of $(s+1)$-dimensions.

We should note that the open $s$-braneoid theory at
the level of action functional see the structure of
$(s+1)$-algebra of   the  cohomology $H^*(\Hoch(A(\L)_s))$ rather than
that of the Hochschild complex $\Hoch(A(\L)_s)$.
There is a fundamental theorem that there is a structure of 
$(s+1)$-algebra
on the Hochschild complex of any $s$-algebra. 
Such is an $(s+1)$-algebra
is called {\it formal} if it is quasi-isomorphic to its cohomology.
The formality means that the set of equivalence class of solutions of
Maurer-Cartan equation \aag\ of the cohomological $(s+1)$-algebra
is isomorphic to the set of equivalence class of solutions of
Maurer-Cartan equation of a $(s+1)$-algebra structure on
$\Hoch(A_s(\L))$.

Assume that $N_s$ bounds an oriented compact $(s+1)$-dimensional
manifold $M_{s+1}$. We may regard a topological open $s$-braneoid
theory on $M_{s+1}$
as closed $(s-1)$-brane theory on $\rd M_{s+1}=N_s$ with 'bulk" 
deformations
specified by $\mH^{s+1}$.
The boundary sector of the theory may be viewed as
the theory maps $\bos{\w}: \rd M_{s+1}\longrightarrow \L$.
Recall a solution of the "master" equation \aag\ induces a structure
of strongly homotopy $(s-1)$-Poisson structure on $\L$
or, equivalently the structure
$(\mf_{\L};(\m_1,\m_2,\m_3,\ldots )_{\L})$
of strongly homotopy Lie $(s-1)$-algebroid or,
simply, of $s$-algebra.
A special case of such structure is a degree $U=s-1$ symplectic structure
on $\L$ or, equivalently, the structure  of
strongly homotopy Lie $(s-1)$-algebroid with $\m_n =0$ for all $n$ 
except for
$n=2$ and $\m_2$ is non-degenerated.
Then we just have the standard topological closed $(s-1)$-brane theory
associated with $\T_{s}=\L$ such that $\{.,.\}_{s-1}\equiv \m_2$.
In general we may use the above topological open $s$-brane on
$M_{s+1}$ to define differential-topological invariants of the boundary
$N_s =M_{s+1}$ by the correlation functions of the boundary
BV observables. In other words topological open $s$-brane associates
any $s$-algebra -strongly homotopy $(s-1)$-algebroid, with 
differential-topological invariants of
$s$-dimensional manifold, which bounds $(s+1)$-dimensional space.

It seems to be reasonable to believe that
the perturbation expansions of the open $s$-brane theory above
generates elements of Hochschild complex  $\Hoch(A(\L)_s)$
and the BV Ward identity of the theory\foot{The BV Ward identity is
an identity of path integral as the result of BV master equations}
is equivalent to the Maurer-Cartan equation for
$\Hoch(A(\L)_s)$.
The above "principle" is beautifully
demonstrated by Cattaneo-Felder for $s=1$ and $\L =X$
is a Poisson manifold in their path integral approach to
Kontsevich's formality theorem \cite{CF}.
We emphasis that the Maurer-Cartan equation for
the cohomological $(s+1)$-algebra
$H^*(\Hoch(A(\L)_s))$ is equivalent to the BV master equation
of our $s$-brane theory and the BV Ward identity is a direct consequence,
at lease formally, of the BV master equation. A crucial point is
that the BV Ward identity depends differential-topology of $M_{s+1}$
(including configurations space of points on  $M_{s+1}$).

The above discussion seem to indicate that they may be fundamental relations 
between differential-topology
of $d$-dimensions and the world of $d$-algebras.
It could be more precise to state that (quantum) $d$-algebra should be
defined in terms of differential-topology of $d$-dimensions as detected
by path integrals.
Kontsevich conjectured that there are structures of
$d$-algebra in conformal field theory on $\R^d$ with
motivic Galois group action on the moduli space \cite{K2}.
Our construction naturally suggest that the conjecture can be naturally
generalized to  $d$-braneoids  on any orient smooth
$(d+1)$-dimensional manifold.  All those direct us to certain universal
properties of Feynman path integrals of the theory related with
differential-topology, arithmetic geometry as well as number theory.

\subsubsection{Multiple Boundaries}

So far we assume that $M_{s+1}$ has only a single boundary
component. Now we relax the condition by allowing
$\rd M_{s+1}$ has multiple components. For each component 
$N_i$ of $\rd M_{s+1}$, we pick a Lagrangian subspace
$\L_i$ in $(\T_{s+1},\o_s)$ and assign boundary condition
prescribed before. Now the classical BV master equations
requires that 
\eqn\dfa{
H|_{\L_i}=0, \quad\hbox{for }\forall i
}
in addition to the condition $[H,H]_{-s}=0$.
Then under  the canonical transformation generated
by $\bos{\Psi}$ in \ddq\ we have
\eqn\dfb{
\bos{S}^{(0)\bos{\Psi}}_H= 
\bos{S}^{(0)}_{H_\Psi} +\sum_i \oint_{T[1]N_i} \G_i,
}
where $\G_i = \Psi|_{\L_i}$. Above action functional
satisfies the master equation if and only if
$H_{\Psi}|_{\L_i} =0$ for $\forall i$.
Thus it is obvious that $\Psi \in \mt_s$ satisfying $\Psi|_{\cup_i\L_i} =0$
leads to the same physical theory. 

Let's consider, as a simplest example, the case that
$\rd M_{s+1}= N_1 \cup N_2$ and the associated
Lagrangian subspaces $\L_1$ and $\L_2$ in $(\T_{s+1},\o_s)$
are complementary
with each others defined by $p_a =0$ and $q^a=0$,respectively, for $\forall a$.
We may expand $H$ as $H= \sum_{k,\ell=1}^\infty\Fr{1}{k! \ell !}m^{a_1\ldots a_k}_{b_1\ldots
b_\ell} q^{b_1}\cdots  q^{b_\ell} p_{a_1}\cdots p_{a_n}$,
and for each $\L_1$ and $\L_2$ we have
structures of sh Lie $s$-algebroid, which, together, may be called
sh Lie $s$-bialgebroids.
Now we have the associated classical BV action functional
\eqn\dfc{
\bos{S}^{(0)}_H 
=\oint_{T[1]M_{s+1}}\left( \hat p_a\hat d\hat q^a 
+ \sum_{k,\ell=1}^\infty\Fr{1}{k! \ell !}m^{a_1\ldots a_k}_{b_1\ldots
b_\ell}\hat q^{b_1}\cdots \hat q^{b_\ell}\hat p_{a_1}\cdots\hat p_{a_n}  \right).
}
Note that the (super) propagator exist between $\hat p_a$ and
$\hat q^a$ only such that at each boundary there are no propagation
and interaction takes place only at the bulk.
The tree-level interactions correspond to the strucuture
of {\it classical} sh Lie $s$-bialgebroids, while the higher order
(in $\hbar$) corrections would lead to {\it quantum} sh Lie $s$-bialgebroids
We may say the QFT defines some kind of quantum cobordism.

\subsection{Toward Quantum Clouds}

This paper has a serious limitation to unveil quantum world. The
general results in section $3$ is based on assumption that
$\bos{\Delta}$ exists without proper definition of it, while using $\bos{\Delta}$
crucially.  In section $5$, where we present realistic case of $d$-dimensional
QFTs and their family, we simply ignored $\bos{\Delta}$, while written in the
quantum perspective  relying on  $\bos{\Delta}$.
Let $M_n$ be a $n$-manifold with or without boundary.
Let $\{\phi(\s)^a, \phi(\s)^\bullet_a\}$ be {\tt fields} and {\tt anti-fields}
of certain model discussed in section $5$ with suitable boundary
conditions, where
$\{\s^\m\}$ is a local coordinates system in $M_n$ and $\{a\}$
denotes all the discrete indices in the model.
Then $\bos{\Delta}$ is naively "defined" as follows
\eqn\xxx{
\bos{\Delta} "="\lim_{\t^\m \rightarrow \s^\m}
\sum_a \int\! \sqrt{g}d^n \t\int\!\sqrt{g}d^n \s\left( (-1)^{|\phi^a|+1}\Fr{\d^2}{\d \phi(\t)^\bullet_a\d \phi(\s)^a}\right)
}
involving diagonal of $M_n\times M_n$ which requires suitable regularization.
Related to above Feynman propagators are certain differential forms on
suitably compactified configuration space of points on $M_n$. Also, due to non-local
observables, supported on various cycles, we need to worry about space
of cycles in $M_n$. The pressing problem is to define 
$\bos{\Delta}$ correctly and universally (before doing any gauge fixing) in each 
dimensions for general smooth manifold with or without boundary (this will take care of
the crucial renormalization of QFT). We remark that a closely related
problem, though in a limited situation (loop space), has been dealt in string topology
of Chase and Sullivan \cite{CS}.
Once this is achieve QFTs are
largely characterized by certain $\bos{\CK_\CS}$-cohomology ring,
for $\bos{\Delta}e^{-\bos{\CS}/\hbar}=0$, as we demonstrated in section $3$.
Then we may concentrate on  structures associated
with (correct) moduli space of QFTs for both generic part and singular points, 
which give rise to various perturbative QFTs.
As for mathematical side, it will give us an universal quantization machine 
for $d$-algebras decorated by algebro-differential-topology of $d/(d+1)$-manifolds.

\bigskip
\baselineskip=10pt 
{\footnotesize

I am grateful to Bumsig Kim for kindly providing me an excellent research environment 
at Math Dept.\ of  POSTECH, as well as to Dosang Joe for their moral supports 
and discussions. I also thanks Jaemo Park for a temporary
position at APCTP during a transition period.

I am grateful to Iouri Chepelev, Christiaan Hofman, Seungjoon Hyun,
Young-Geun Oh, John Terrila for collaborations in
various closely related subjects.
The some results here and in other
related works, yet to be published, has been announced in several
occasions in YITP and math departments of Stony Brook, Penn State,
CUNY grad. center, Northwestern, Kyungpook,
Wisconsin-Madison and Physics. Dept of Younsei,
at Kyungjoo Conf. on Frobenius structure and
singularity theory, RIMS and KIAS. I would like to thanks  my hosts Martin Rocek, Dimitry Roytenberg,   
Dennis Sullivan, Ezra Getzler, Hoil Kim, Yong-Geun Oh, Seungjoon Hyun, Bumsig Kim,
Koiji Saito and Piljin Yi for their hospitality and discussions. 
I am grateful to  Dennis Sullivan  for many stimulating
discussions on quantization of anything and to Jim Stasheff for useful
communications and comments.
}

\end{document}